\documentclass[a4paper,11pt]{article}
\pdfoutput=1 

\usepackage{jheppub} 
\usepackage{bbm}
\usepackage{xcolor}
\usepackage[T1]{fontenc} 
 
\begin{document} 

\title{Phases of Surface Defects in Scalar Field Theories}

\author[a]{Avia Raviv-Moshe}                       
\author[a,b]{and Siwei Zhong}

              \affiliation[a]{Simons Center for Geometry and Physics, SUNY, Stony Brook, NY 11794, USA}      
              \affiliation[b]{C. N. Yang Institute for Theoretical Physics, Stony Brook University, Stony Brook, NY 11794, USA}

          \emailAdd{araviv-moshe@scgp.stonybrook.edu}
                    \emailAdd{siwei.zhong@stonybrook.edu}

\abstract{
We study mass-type surface defects in a free scalar and Wilson-Fisher (WF) $O(N)$ theories. We obtain exact results for the free scalar defect, including its RG
flow and defect Weyl anomaly. We classify phases of such defects at the WF fixed point near four
dimensions, whose perturbative RG flow is investigated. We propose an IR effective action for the
non-perturbative regime and check its self-consistency.
}

\maketitle
\flushbottom

\section{Introduction and Summary}

The study of defects and boundaries in Conformal Field Theories (CFTs) has attracted much attention in recent years. As the study of point defects (local operators) in CFTs revealed a rich physical structure that rendered many insights, extended defects provide a refined understanding of quantum field theories. For example, symmetries can be phrased in topological defects \cite{gaiotto2015generalized,roumpedakis2022higher} and Wilson lines probe phases of gauge theories \cite{gaiotto2015generalized,aharony2022phases}. In this work, we will focus on two-dimensional surface defects in scalar CFTs of dimension $d$. We note that surface defects exhibit rich physics in different theories, for example, in 4d $\mathcal{N}=4$ SYM \cite{gukov2006gauge,drukker2008probing,gukov2010rigid, Wang:2020seq}, in 4d Maxwell theory \cite{Herzog:2022jqv}, in scalar CFTs similar to the subjects of this paper \cite{lauria2021line,krishnan2023plane}, and in the context of conformal boundaries for 3d theories \cite{burkhardt1987surface,mcavity1995conformal,ohno19831, dimofte2018dual,metlitski2022boundary}. We shall elaborate on terminologies in the following.

In Euclidean signature, the conformal symmetry of the background theory is $SO(d+1,1)$, which is explicitly broken by the presence of a defect. The term {\it Defect Conformal Field Theory} (DCFT) refers to the case where the system preserves the maximal conformal subgroup. In this paper, we will mainly consider a plane defect $\mathbb{R}^2$ embedded in flat space $\mathbb{R}^d$, where DCFTs are of the symmetry $SO(3,1)\times SO(d-2)$. Starting from a UV DCFT, one could add relevant perturbations to the defect action and trigger a {\it Renormalization Group} (RG) flow. Schematically, it can be described by:

\begin{equation} \label{eq_rg_surface}
S_{\text{DCFT}} \to S_{\text{DCFT}} + \Lambda^{2-\Delta_{\mathcal{O}}} \int_{\Sigma} d^2 \sigma\, \mathcal{O}(\sigma)\, , 
\end{equation}
where $d^2\sigma$ is the diff-invariant integration on the surface $\Sigma$ associated with the defect, $\Lambda$ is the UV-scale, and $\mathcal{O}$ is an operator in the UV DCFT with $\Delta_\mathcal{O}\leq 2$. Generally (but not always\footnote{Other possibilities include, for example, the runaway behavior \cite{Cuomo:2022xgw}. }), the defect RG flow will end at an IR DCFT. We will apply such a method of defect construction throughout this paper, by adding mass-type deformations to the trivial DCFT and investigating its defect RG flow fixed point.  

In unitary and local theories, an important theorem concerning the RG flow on surface defect is that there exists a $b$-coefficient that satisfies $b_{\text{UV}} \geq b_{\text{IR}}$ \cite{jensen2016constraint} (see also \cite{wang2021surface,shachar2022rg}). Such a $b$-coefficient is defined by the defect Weyl anomaly. Explicitly, for an infinitesimal Weyl variation $\omega$,\footnote{With the background metric $g_{\mu\nu}$, the Weyl variation is $\delta_\omega g_{\mu\nu} = 2\omega g_{\mu\nu}$. Generally, the Weyl anomaly $\mathcal{A}=\mathcal{A}_\text{b}+\delta^{d-2}(\Sigma)\mathcal{A}_\text{d}$ consists a background and a defect contribution. 
 The $b$-coefficient is defined as $\mathcal{A}_d = \frac{b}{24\pi}\hat{R}+\cdots$, where the dots stand for other (independent) geometric contributions to the defect Weyl anomaly.} the partition function $Z$ changes as \cite{henningson1999weyl,schwimmer2008entanglement,graham1999conformal}

\begin{equation}\label{eq_anomaly_def_intro}
\delta_\omega\left(\log Z\right)=\frac{b}{24\pi}\int_{\Sigma}d^2\sigma \omega(\sigma) \hat{R}(\sigma)+\cdots
\end{equation}
where $\hat{R}$ is the Ricci scalar curvature of the induced metric on the defect, and we have omitted other geometric contributions. As an observable for DCFTs,  the $b$-coefficient provides a useful tool for the study of RG flows on surface defects, as will be discussed through this paper.

Below we briefly summarize our main results:
\begin{itemize}
\item The mass-type surface defect in a free scalar theory is solved exactly. We find that in dimensions $2\leq d \leq 4$ the theory admits a single IR-stable fixed point, whose special cases are discussed:
\begin{itemize}
\item At $d=4$ it coincides with the trivial one.
\item At $d=3$ it represents two copies of the free field Dirichlet boundary conditions.
\item At $d=2$ it reproduces a trivially gapped theory.
\end{itemize}
An exact $b$-coefficient is calculated for such a fixed point, and we find the result $b=-(2-d/2)^3$. Our general result agrees with the perturbative calculation near four dimensions \cite{shachar2022rg}, and for $d=3$ it reduces to the result found in \cite{jensen2016constraint}. 

\item The phase diagram of a mass-type surface defect \footnote{By mass-type here we refer to deformations that preserve the $\mathbb{Z}_2\subseteq O(N)$. In critical spin-lattice realization of the WF fixed point, the $\mathbb{Z}_2$ is the spin-flip symmetry.} in the $O(N)$ Wilson-Fisher fixed point near four-dimension is studied. We analyze the defect RG flows both within and outside of the perturbative regime:

\begin{itemize}
\item In the perturbative regime we calculate the beta-functions associated with the defect RG flow.  At the one-loop level, we find different phase diagrams for $N<6$, $6<N<10$, and $N>10$ (illustrated in figure \ref{Phase Diagram}).  Theories with $N=6$ and $N=10$ in some subcases land on similar classifications, while other subcases depend on higher orders in perturbation theory. 

\item Outside the perturbative regime we propose an IR-effective action (given in equation \eqref{IR effective}) and verify its self-consistency. We study the mean-field saddle point and obtain conformal data by mapping and solving the theory in a Euclidean $AdS$ space.
\end{itemize}
\end{itemize}
To make contact with previous studies, the large-$N$ limit of the 3d Wilson-Fisher fixed point in the presence of $O(N)$-preserving mass-type surface defect is investigated in \cite{krishnan2023plane}, while general numerical studies were conducted in \cite{PhysRevE.72.016128,PhysRevE.73.056116,ParisenToldin:2020gpb,Hu:2021xdy,Toldin:2021kun,Padayasi:2021sik}.

This paper is organized as follows. In section \ref{sec_Free_Theory}, we study 
the mass-type surface defect in a free scalar background theory. In section \ref{sec_TheON_critical_perturbationTheory}, we study mass-type surface defects in $O(N)$ Wilson-Fisher fixed point near four-dimension and investigate perturbative DCFTs. In section \ref{sec_Effective_Action}, we continue the discussion in section \ref{sec_TheON_critical_perturbationTheory} and propose IR-effective theories for DCFTs outside the perturbative regime.

\paragraph{Note added:} While we were completing this work,  we became aware of upcoming papers \cite{2023Giombi} and \cite{Trepanier:2023tvb}, which present results that overlap with parts of this work.  
We are grateful to the authors of \cite{2023Giombi} for sharing a preliminary version of their draft and coordinating the submission date.

\section{The Free Theory: A Solvable Model}
\label{sec_Free_Theory}
We start by considering perhaps the simplest possible model: adding a surface defect localized on a flat two-dimensional plane to a background theory consisting of a single free scalar field. Such a defect is constructed using the background's physical degrees of freedom and is taken to be quadratic in the scalar field. The quadratic operator is classically marginal on the defect with the background dimension $d=4$ and becomes relevant when $d<4$.

The action in Euclidean signature reads :
\begin{equation}
\label{exactly solvable}
\begin{aligned}
S=\frac{1}{2}\int_{\mathbb{R}^{d}}d^{d}x(\partial \phi)^2+\frac{\gamma_{\text{b}}}{2}\int_{\mathbb{R}^2}d^2z\,\phi^2 ,
    \end{aligned}
\end{equation}
where $\phi$ is a real scalar field and $\gamma_b$ is the bare defect coupling constant. Here and throughout this paper, we use $z$ to denote the coordinate system of the defect ($z=(z^1,z^2)$). Coordinates associated with the $\mathbb{R}^d$, in which the background theory is defined, are denoted by $x=x^\mu=(y,z)$, $\mu = 1, \cdots, d$. The defect is located at $(y=0,z)$, where $y$ is used to denote the orthogonal directions. In the following,  we will also denote $d=4-\epsilon$,  with $\epsilon\geq 0$ a dimensionless parameter. Note that analysis in this section will not require the perturbative condition $\epsilon \ll 1$

The model \eqref{exactly solvable} is Gaussian and hence can be solved exactly. In the following two subsections, we solve the theory for $2\leq d\leq 4$ and calculate the exact beta function associated with the dimensionless renormalized defect coupling. A stable IR fixed point is found, and we calculate the $b$-coefficient \cite{jensen2016constraint}, the coefficient of the Euler density in the defect's Weyl anomaly, at such a fixed point.

Before proceeding to calculations, we comment that one could consider a generalization of $N$ free scalar fields with an action given by:
\begin{equation}
\label{O(N) analgog}
\begin{aligned}
S=\int_{\mathbb{R}^d}d^dx\,\frac{1}{2}(\partial \phi^i)^2+\frac{1}{2}\int_{\mathbb{R}^2}d^2z(\gamma_b)_{ij}\phi^i\phi^j,
    \end{aligned}
\end{equation}
where $\{i,j\}=1,\cdots, N$, and $\{(\gamma_b)_{ij}\}$ is the bare coupling tensor. However, in contrast to the interacting theory discussed in the next section, a field redefinition can be facilitated to diagonalize the defect action. Hence, this problem is completely equivalent to considering ($N$ independent copies of) the model \eqref{exactly solvable}.

\subsection{Exact RG}

In order to solve the model \eqref{exactly solvable}, we apply the Wilsonian exact renormalization group analysis \cite{wilson1974renormalization}. In such a picture, a UV-observational cutoff $\Lambda$ is introduced, below which we can recast the action \eqref{exactly solvable} in terms of Fourier modes as:
\begin{equation}
\begin{aligned}
S(\Lambda)=\frac{1}{2}\sum_{k_{1,2}^2+p_{1,2}^2\leq \Lambda^2 }\left \{ (k_1^2+p_{1}^2)\delta_{k_1+k_2}\delta_{p_{1}+p_{2}}+\gamma_{\text{b}}\delta_{k_1+k_2}\right \} \phi_{k_1,p_{1}}\phi_{k_2,p_{2}},
    \end{aligned}
\end{equation}
where we have let $(p)$ be the momentum vector conjugate to coordinate $(y)$, and $(k)$ be that conjugate to $(z)$.

Corresponding to the coarse-graining procedure, we lower the cutoff $\Lambda$ to ${\Lambda}'$ and integrate the modes between them to obtain an effective theory describing IR physics. Without the defect, the theory is free and Poincare symmetry is preserved. Therefore, in that case, UV modes decouple from the IR modes and there is no renormalization aside from the trivial scaling behavior. However, in the presence of a defect, UV modes couple linearly to IR modes, and the difference in the actions $S(\Lambda)-S(\Lambda')$ reads:
\begin{equation}
\begin{aligned}
S(\Lambda)-S({\Lambda}')\sim\sum_{k^2\leq {\Lambda}'^2}\left\{ \frac{1}{2}{\sum_{p_{1}}}'{\sum_{p_{2}}}'\left [\delta_{p_{1}+p_{2}}(k^2+p_{1}^2)+\gamma_{\text{b}}\right ]\phi_{-k,p_1}\phi_{k,p_2}\right .
\\ \left. +\gamma_{\text{b}}\left[ \sum_{p^2\leq {\Lambda}'^2-k^2}\phi_{-k,p}\right]\left[ {\sum_{p_1}}'\phi_{k,p_{1}}\right]\right \} \, ,
    \end{aligned}
\end{equation}
where we use the notation ${\sum}'$ to denote a summation over the momentum shell of ${\Lambda}'^2-k^2\leq p_{1,2}^2\leq \Lambda^2-k^2$. The above yields a standard Gaussian integral, and by noticing that the inverse matrix is:
\begin{equation}
\label{propagator}
\begin{aligned}
\left [\delta_{p_1+p_2}(k^2+p_1^2)+\gamma_{\text{b}}\right ]^{-1}=\frac{\delta_{p_1+p_2}}{k^2+p_{1}^2}-\frac{\gamma_{\text{b}}}{(k^2+p_1^2)(k^2+p_2^2)(1+\gamma_{\text{b}}{\sum}'\frac{1}{k^2+p^2})}\,,
    \end{aligned}
\end{equation}
we can read the Wilsonian effective term for IR modes:
\begin{equation}
\label{eq_effective_action1}
\begin{aligned}
S_{\text{eff}}=&-\frac{\gamma_{\text{b}}}{2}\sum_{k_{1,2}^2+p_{1,2}^2\leq \Lambda^2 }\left(1-\frac{1}{1+\gamma_{\text{b}}{\sum}'\frac{1}{k_1^2+p^2}}\right)\delta_{k_1+k_2}\phi_{k_1,p_1}\phi_{k_2,p_2}.
    \end{aligned}
\end{equation}
The summation in the denominator can be evaluated explicitly:
\begin{equation}
\label{eq_integral1}
\begin{aligned}
{\sum_{p}}'\frac{1}{k^2+p^2}=\int_{{\Lambda}'^2-k^2\leq p^2\leq \Lambda^2-k^2}\frac{d^{2-\epsilon}p}{(2\pi)^{2-\epsilon}}\frac{1}{k^2+p^2}=\frac{2^{\epsilon } \left(\Lambda^{-\epsilon
   }-{\Lambda}'^{-\epsilon
   }\right)}{ \pi ^{1-\frac{\epsilon }{2}} \epsilon ^2 \Gamma \left(-\frac{\epsilon }{2}\right)}+O(k^2)\,.
    \end{aligned}
\end{equation}
In the above equation, terms of order $O(k^2)$ stand for kinematic terms, dynamically generated at the surface defect, which are manifestly irrelevant (these can in principle be calculated but will depend on the regulation scheme).

We define $\eta\equiv-\log({\Lambda}'/\Lambda)\geq 0$ the position along the RG flow, and the renormalized coupling $\gamma\equiv{\Lambda}^{-\epsilon}\gamma_{\text{b}}$. From the Wilsonian effective term \eqref{eq_effective_action1}, using the integral in equation \eqref{eq_integral1}, the exact RG running of the dimensionless coupling $\gamma$ reads:

\begin{equation}
\label{exact rg}
\begin{aligned}
\frac{1}{\gamma(\eta)}=\frac{(1-e^{-\epsilon \eta})}{2^{1-\epsilon}\pi^{1-\frac{\epsilon}{2}}\epsilon\Gamma(1-\frac{\epsilon}{2})}+\frac{e^{-\epsilon \eta}}{\gamma(0)}.
    \end{aligned}
\end{equation}
which corresponds to a one-loop exact defect RG flow. \footnote{Similar results can be derived by mapping the theory to $\mathbb{H}^3\times S^{d-3}$ and studying the corresponding boundary conditions.} From the result in \eqref{exact rg}, $\gamma>0$ deformation of the trivial DCFT triggers a flow to an IR-stable interface, with a fixed point value:
\begin{equation}
\label{eq_fixed_point}
\gamma=\gamma_*\equiv 2^{1-\epsilon}\pi^{1-\frac{\epsilon}{2}}\epsilon\, \Gamma(1-\frac{\epsilon}{2}).
\end{equation}
For $\epsilon \ll 1$, the above reads $\gamma_* = 2\pi \epsilon + O(\epsilon^2)$, in agreement with the perturbative result found in \cite{shachar2022rg}.

\begin{figure}[!h]
\centering
  \includegraphics[width=0.6\textwidth ]{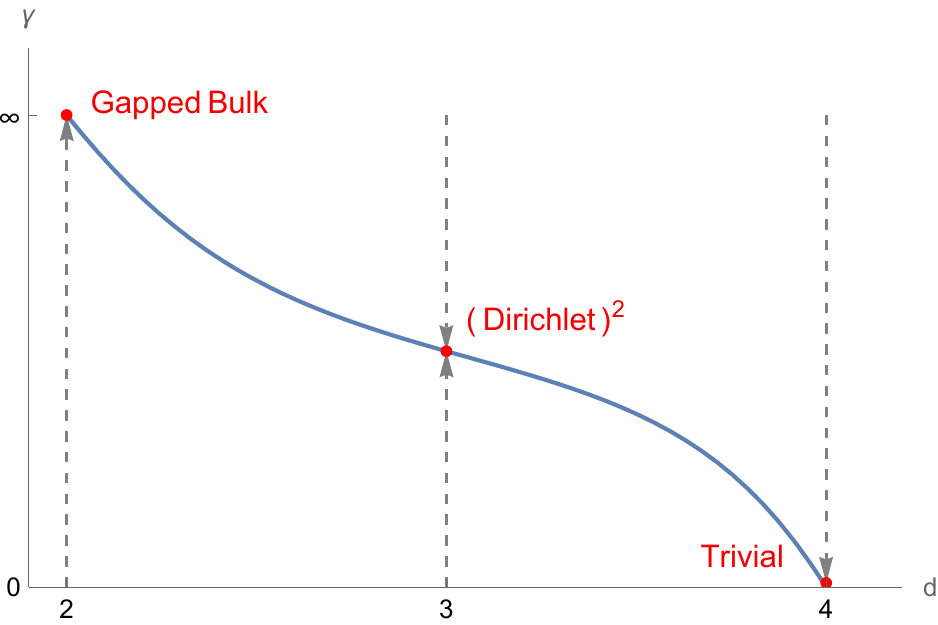}
  \caption{\label{Free FP}IR-stable fixed points in the free theory. The blue curve represents the fixed point dependence on the background theory dimension, and the grey dashed arrow indicates the RG flow from UV to IR. Fixed points with physical interpretations in integer dimensions are marked in red. }
\end{figure}

In a four-dimensional background $d=4$,  equation \eqref{exact rg} becomes:
\begin{equation}
\begin{aligned}
\frac{1}{\gamma(\eta)}=\frac{\eta}{2\pi}+\frac{1}{\gamma(0)},
    \end{aligned}
\end{equation}
indicating such a $\gamma>0$ defect deformation is irrelevant, in agreement with the general analysis found in \cite{lauria2021line}. When $d=3$, we find $\gamma_*=\pi$ in this renormalization scheme. When $d=2$, $\gamma_*=\infty$ and equation \eqref{exact rg} becomes the trivial scaling of the scalar mass operator. We will elaborate on the fixed point physical meanings and subtleties in the next subsection, and here we get ahead of ourselves and present the phase diagram \ref{Free FP}.

Lastly, we comment on an RG flow triggered by a $\gamma<0$ deformation. On the one hand, in this case equation \eqref{exact rg} does not end at a finite fixed point. Instead, there exists a finite scale in which the 1-loop exactness of RG breaks down. On the other hand, we point out that the Hamiltonian is not bounded from below when $\gamma<0$ for a free background theory, and we expect that in this case there is a runaway behavior \footnote{We conjecture that for $\gamma<0$,  the b-coefficient  (see equation \eqref{eq_defect_partition_func_contr}) in the IR is unbounded from below. This resembles the case of pinning field line defect in a free scalar theory \cite{cuomo2022localized}. There, the defect entropy $s$ \cite{Cuomo:2021rkm} satisfies $s_{\text{IR}}\to -\infty$. }. However, for an interacting theory whose background potential is bounded from below, it could be the case where such a defect induces localized degrees of freedom and flows to a healthy DCFT in the IR. Such a setup will be studied in more detail in sections \ref{sec_TheON_critical_perturbationTheory} and \ref{sec_Effective_Action}. 

\subsection{Defect Weyl Anomaly}

 In this subsection, we calculate the $b$-coefficient at the IR stable fixed point \eqref{eq_fixed_point} and investigate its physics. Before the calculation, we notice that since the DCFT is still Gaussian and can be analyzed through Wick contraction, it falls into the category of a generalized free theory. The lowest-lying nontrivial defect operator $\hat\phi$ has the scaling dimension 
\begin{equation}
\begin{aligned}
\Delta(\hat\phi)=1+\frac{\epsilon}{2},
    \end{aligned}
\end{equation}
at the stable IR fixed point \eqref{eq_fixed_point}. 
This bears similarities to double-trace deformations in a large-N two-dimensional CFT \cite{gubser2003universal}, which motivates us to study the Weyl anomaly b-coefficient through the defect contribution to the free energy $\mathcal{F}$ \cite{jensen2016constraint,shachar2022rg}.

Consider the defect \eqref{exactly solvable} of spherical geometry, that is, a $\mathbb{S}^2$ sphere of radius $R$ embedded in $\mathbb{R}^d$. The defect contribution to the free energy is defined by:
\begin{equation}\label{eq_defect_partition_func_contr}
\mathcal{F}\equiv-\left(\log Z^{\text{DCFT}} - \log Z^{\text{CFT}}\right),
\end{equation}
where $Z^{\text{DCFT}}$ is the partition function of the full theory with the presence of the defect, and $Z^{\text{CFT}}$ is that of the same background theory but without the defect. $\mathcal{F}$ generally depends on the regulation scheme and suffers from ambiguities. However, the $b$-coefficient, which can be extracted from the logarithmic IR-divergence of \eqref{eq_defect_partition_func_contr}, is universal and scheme-independent \cite{jensen2016constraint,shachar2022rg}. In the $\mathbb{S}^2$ defect geometry, we have :
\begin{equation}
\label{b-def}
\begin{aligned}
\mathcal{F}\sim a_0 +a_1 (\Lambda R)^2+ \frac{b}{3}\log (\Lambda R),
    \end{aligned}
\end{equation}
where $a_1$ and $a_0$ are non-universal coefficients and $b$ satisfies the inequality $b_{\text{UV}}\geq b_{\text{IR}}$ at the UV and IR fixed points respectively \cite{jensen2016constraint,shachar2022rg}. Note that such a statement is for fixed point theories, and to avoid subtleties in the middle of the RG flow we will set the bare coupling to be at the stable IR fixed point \eqref{eq_fixed_point} in the following, which reads:
\begin{equation}
\begin{aligned}
(\gamma_{\text{b}})_*=2 \pi \epsilon\Gamma(1-\frac{\epsilon}{2}) \left(\frac{\Lambda}{2 \sqrt{\pi}}\right)^{\epsilon}.
    \end{aligned}
\end{equation}

To derive $\mathcal{F}$, we will need to investigate the Laplacian operator eigenvalues in the presence of the defect. Due to the defect isometry group $SO(3)$, the Laplacian is diagonal in the momentum space dual to the $\mathbb{S}^2$ sphere coordinate. The free field propagator of spherical harmonic waves $\{\alpha_l\}$ (see equation \eqref{eq_app4} for definition) are given by:
\begin{equation}
\label{al def}
\begin{aligned}
\alpha_l(\epsilon)\equiv \frac{\left(\pi \epsilon/2\right)   \Gamma \left(l-\frac{\epsilon }{2}+1\right)}{\sin \left(\pi \epsilon/2\right)\Gamma \left(l+\frac{\epsilon
   }{2}+1\right)}. 
    \end{aligned}
\end{equation}
For a Gaussian theory as \eqref{exactly solvable}, $\mathcal{F}$ can be evaluated using basic linear algebra (see appendix \ref{Regulation} for details). It reads: 
\begin{equation}
\label{F-function}
\begin{aligned}
\mathcal{F}=\frac{1}{2}\sum_{l\geq0}(2l+1)\log{\left(1+ (\Lambda R)^{\epsilon} \alpha_l(\epsilon) \right)},
    \end{aligned}
\end{equation}
which diverges in the IR limit $ R \rightarrow \infty$. To extract the IR information from the above expression we will apply a dimensional regulation scheme (similar to the analysis found in  \cite{klebanov2011f,diaz2007partition}) on the defect dimension $\Tilde{d}$ (instead of the background space). In the IR limit, equation \eqref{F-function} can be simplified as follows (see e.g. \cite{diaz2007partition}):
\begin{equation}
\label{F-function IR limit}
\begin{aligned}
\mathcal{F}=-\frac{1}{2}\sum_{l\geq0}\text{deg}(\tilde{d},l)\log{\frac{ \Gamma \left(l+\frac{\Tilde{d}}{2} +\frac{\epsilon }{2}\right)}{ \Gamma \left(l+\frac{\Tilde{d}}{2}-\frac{\epsilon }{2}\right)}} \, ,
    \end{aligned}
\end{equation}
where 
\begin{equation}\label{eq_harmonics_degeneracy}
\text{deg}(\tilde{d},l)\equiv (2l+\Tilde{d}-1)\frac{\Gamma (l+\Tilde{d} -1)}{\Gamma (l+1)\Gamma (\Tilde{d})},
\end{equation}
stands for the spherical harmonic degeneracy.  
In this regularization scheme, physical quantities are obtained by an analytic continuation to $\Tilde{d}\rightarrow 2$. As in ordinary even dimensional CFTs, such a $\mathcal{F}$-function exhibits IR-divergence in $\Tilde{\epsilon}\equiv \Tilde{d}-2$. Thus, in this case, the pole in $O(1/\Tilde{\epsilon})$ corresponds to the logtharmic IR-divergence in the cutoff regulation parameter (see e.g. \cite{graham1999volume,diaz2007partition}), and shall be identified as the defect Weyl anomaly coefficient in equation \eqref{eq_anomaly_def_intro}. We find ($d\equiv 4-\epsilon$):
\begin{equation}\label{eq_defect_central_charge_free}
\begin{aligned}
b_{\text{IR}}=-\frac{\epsilon^3}{8}.
    \end{aligned}
\end{equation}

The first consistency check is $b_{\text{IR}}<0$ when $d<4$. Note that at the UV fixed point of \ref{exact rg}, the DCFT is trivial with $b_{\text{UV}}=0$. Hence indeed the inequality  $b_{\text{UV}}\geq b_{\text{IR}}$ is satisfied in this example. The second check is when $\epsilon \ll 1$, \eqref{eq_defect_central_charge_free} agrees with the perturbative result in \cite{shachar2022rg}, and there are no higher order corrections. We comment that this is a consequence of the defect RG being 1-loop exact.

At $d=3$, the $b$-coefficient takes the value of two copies of the free field Dirichlet boundary condition ($b_{\text{D}}=-1/16$) \cite{jensen2016constraint}. This observation is supported by the propagator at the fixed point $(\gamma_b)_*=\pi \Lambda$:

\begin{equation}
\begin{aligned}
\left [\delta_{p_1+p_2}(k^2+p_1^2)+\pi \Lambda\right ]^{-1}=
\frac{\delta_{p_1+p_2}}{k^2+p_1^2}-\frac{2|k|}{(k^2+p_1^2)(k^2+p_2^2)}+O\left(\Lambda^{-3}\right).
    \end{aligned}
\end{equation}
where the subleading terms are suppressed by the cutoff $\Lambda$ and depend on the regulation scheme we chose. To see its physical meaning, we evaluate the layer susceptibility \cite{dey2020operator,shpot2021boundary}

\begin{equation}
\begin{aligned}
\chi(y_1,y_2):=&\lim_{|k|\rightarrow 0}\int \frac{d p_1}{2\pi} \frac{d p_2}{2\pi} e^{i (p_1 y_1+p_2 y_2)}\left(\frac{\delta_{p_1+p_2}}{k^2+p_1^2}-\frac{2|k|}{(k^2+p_1^2)(k^2+p_2^2)}\right)\\
=&\lim_{|k|\rightarrow 0}\left(\frac{e^{-|k||y_1-y_2|}}{2|k|}-\frac{e^{-|k|(|y_1|+|y_2|)}}{2|k|}\right)\\
=&\frac{1}{2}(|y_1|+|y_2|-|y_1-y_2|). 
    \end{aligned}
\end{equation}
such that if $y_1$ and $y_2$ are on different sides of the plane defect, the susceptibility vanishes in the IR. In this case, the presence of the defect simply breaks space into two. That is, the theory in $\mathbb{R}^3$ at IR is broken into two distinct copies of free scalar in half-spaces $\mathbb{R}^3_+$ with Dirichlet boundary conditions, which agrees with the result \eqref{eq_defect_central_charge_free} for the $b$-coefficient.

At $d=2$, there is a subtlety since one can no longer perform the Weyl transformation individually to the defect, as it is now indistinguishable from the background theory. It is known that a 2d free scalar (should be thought of as the large radius limit of a compact boson) has a central charge $c=1$ \cite{ginsparg1988applied}. Therefore, the physical Weyl anomaly coefficient is $b_{\text{IR}}+c=0$. We note this agrees with equation \eqref{exact rg} and the phase diagram \ref{Free FP}, such that the fixed point is a two-dimensional gapped trivial theory.

\section{The $O(N)$ Wilson-Fisher Fixed Point  }
\label{sec_TheON_critical_perturbationTheory}
In this section, we study the analog of \eqref{O(N) analgog} with the background theory being  the $O(N)$ interacting model tuned to the Wilson-Fisher fixed point \cite{wilson1974renormalization} in $d=4-\epsilon$ dimension. Throughout this section, we will assume that $\epsilon \ll 1$ is the smallest parameter in the theory and that the defect couplings are of the order $O(\epsilon)$ such that standard perturbation theory is valid. 

The background theory can be described by $N$ scalar fields $\{\phi^i\}$ with $1\leq i \leq N$ coupled through $\phi^4$-interactions in flat space $\mathbb{R}^d$. We will denote the interactions by a fully symmetric (dimensionless) coupling tensor $\{\lambda_{ijkl}\}$, such that: 
\begin{equation}
\label{bulk model}
\begin{aligned}
S_{0}=\int_{\mathbb{R}^d}d^dx\left\{\frac{1}{2}(\partial \phi^i)^2+\frac{\Lambda^{4-d}}{4!}\lambda_{ijkl}\phi^i\phi^j\phi^k\phi^l\right\},
    \end{aligned}
\end{equation}
where $\Lambda$ is the UV scale. The interacting RG fixed point of our interest to its 1-loop level value is given by \cite{wilson1974renormalization}:

\begin{equation}
\label{phi 4 interaction}
\begin{aligned}
(\lambda_{ijkl})_*&= \frac{16\pi^2\epsilon}{N+8}(\delta_{ij}\delta_{kl}+\delta_{ik}\delta_{jl}+\delta_{il}\delta_{jk})+O\left(\epsilon^2\right).
    \end{aligned}
\end{equation}
An implicit assumption in the following discussion is locality: in the defect's presence, we will continue to work in the background fixed point defined by \eqref{phi 4 interaction}. We will use the same coordinate system as in the last section, such that $x=(y,z)$. When an operator is inserted at $(y,0)$ or $(0,z)$ we will abuse the notation to denote them as $(z)$ and $(y)$, correspondingly. Defect perturbations to the background theory considered in this section can be roughly summarized as $\phi$-quadric operators being inserted at the $y=0$ plane, such that the flip symmetry $\mathbb{Z}_2:\phi^i\rightarrow -\phi^i$ is always preserved. 

In the following, we first review in subsection \ref{subsec_background_data} the conformal data of the background theory that will be used in the defect analysis. In subsection \ref{subsec_Defect RG Flow} we perform the perturbative analysis in the theory with defect insertion and obtain the $O(N)$-DCFT fixed points.  Last, in subsection \ref{subsec_DCFT_FP_PER} we analyze in greater detail the fixed point structure upon explicit forms of $O(N)$ breaking deformations that appear in the one-loop level. 
\subsection{Background Data}
\label{subsec_background_data}
At the background fixed point \eqref{phi 4 interaction}, conformally well-defined $\phi$-quadratic operators are packed in irreducible representations of $O(N)$. The lowest-lying $O(N)$-singlet and $O(N)$-symmetric traceless operators are given by \cite{kehrein1995spectrum,henriksson2023critical}:

\begin{equation}
\begin{aligned}
\mathcal{O}(x)=&\frac{1}{2N}\left[\sum_{1\leq k\leq N}(\phi^k)^2\right](x)\,,
    \end{aligned}
\end{equation}
\begin{equation}
\begin{aligned}
T^{ij}(x)=&\left[\phi^i \phi^j-\frac{\delta_{ij}}{N}\sum_{1\leq k\leq N}(\phi^k)^2\right](x)\,.
\end{aligned}
\end{equation}
The two-point functions read \cite{cuomo2022localized}:

\begin{equation}
\begin{aligned}
\langle\mathcal{O} (x_1)\mathcal{O}(x_2)\rangle=&\frac{\mathcal{N}_\mathcal{O}^2}{|x_1-x_2|^{2\Delta_\mathcal{O}}}\,,
    \end{aligned}
\end{equation}
\begin{equation}
\begin{aligned}
 \langle T^{ij}(x_1) T^{kl}(x_2)\rangle =& \left(\delta^{ik}\delta^{jl}+\delta^{il}\delta^{jk}-\frac{2}{N}\delta^{ij}\delta^{kl} \right)\frac{\mathcal{N}_T^2}{|x_1-x_2 |^{2\Delta_T}}.
\end{aligned}
\end{equation}
where:
\begin{equation}
\begin{aligned}
&\Delta_\mathcal{O}=2-\frac{6}{N+8}\epsilon+O\left (\epsilon^2\right);\quad \mathcal{N}_\mathcal{O}=\frac{1}{4\pi^2\sqrt{2N}}+O\left (\epsilon\right)\,,\\
    \end{aligned}
\end{equation}
\begin{equation}
\begin{aligned}
&\Delta_T=2-\frac{N+6}{N+8}\epsilon+O\left (\epsilon^2\right); \quad \mathcal{N}_T=\frac{1}{4\pi^2}+O\left (\epsilon\right)\,.\\
    \end{aligned}
\end{equation}
In what follows, we will also need the three-point correlation functions involving these operators, for which we take the convention 
\begin{equation}
\label{CFT convention}
\begin{aligned}
\langle\mathcal{O}_a (x_1)\mathcal{O}_b(x_2)\mathcal{O}_c (x_3)\rangle&=\frac{C_{\mathcal{O}_a\mathcal{O}_b\mathcal{O}_c}}{|x_1-x_2|^{\Delta_a+\Delta_b-\Delta_c}|x_2-x_3|^{\Delta_b+\Delta_c-\Delta_a}|x_3-x_1|^{\Delta_c+\Delta_a-\Delta_b}}\,,
    \end{aligned}
\end{equation}
where $\mathcal{O}_a=\{ \mathcal{O}(x), T^{ij}(x)\}$. The OPE coefficients $C_{\mathcal{O}_a\mathcal{O}_b\mathcal{O}_c}$ can be perturbatively calculated, and to their leading order in $O(\epsilon^0)$ it yields: 
\begin{align}
C_{\mathcal{O}\mathcal{O}\mathcal{O}}&=\frac{1}{64N^2\pi^6}\,,\\
  C_{\mathcal{O}\mathcal{O}T^{ij}}&=0\,,\\
C_{\mathcal{O}T^{ij}T^{kl}}&=\frac{1}{32N\pi^6}\left(\delta^{ik}\delta^{jl}+\delta^{il}\delta^{jk}-\frac{2}{N}\delta^{ij}\delta^{kl}\right)\,,\\
C_{T^{ij}T^{kl}T^{mn}}&=\frac{1}{64\pi^6}\left(\delta^{ik}\delta^{ln}\delta^{mj}+\delta^{jk}\delta^{ln}\delta^{mi}+\delta^{il}\delta^{kn}\delta^{mj}+\delta^{ik}\delta^{lm}\delta^{nj}\right.\\
+&\delta^{jl}\delta^{km}\delta^{ni}+\delta^{il}\delta^{km}\delta^{nj}+\delta^{jk}\delta^{lm}\delta^{ni}+\delta^{jl}\delta^{kn}\delta^{mi}-\frac{4\delta^{ij}}{N}(\delta^{mk}\delta^{nl}+\delta^{nk}\delta^{ml})\nonumber \\
-&\frac{4\delta^{kl}}{N}(\delta^{im}\delta^{jn}+\delta^{in}\delta^{jm})\left.-\frac{4\delta^{mn}}{N}(\delta^{ik}\delta^{jl}+\delta^{il}\delta^{jk})+\frac{16}{N^2}\delta^{ij}\delta^{kl}\delta^{mn}\right). \nonumber
\end{align}

\subsection{Defect RG Flow}
\label{subsec_Defect RG Flow}
In analogy with equation \eqref{O(N) analgog}, the defect is parametrized by a scalar $\gamma$ and a tensor $g_{ij}$ dimensionless coupling, such that the action is given by:
\begin{equation}\label{eq_action_with_def}
\begin{aligned}
S=S_0+\int_{\mathbb{R}^2}d^2z \Lambda^2 \left\{\frac{\gamma}{\Lambda^{\Delta_\mathcal{O}}} \mathcal{O}+\frac{ g_{ij}}{\Lambda^{\Delta_T}} T^{ij}\right\}\,,
    \end{aligned}
\end{equation}
where $S_0$ is given by equation \eqref{bulk model}, and the background theory is considered at Wilson-Fisher fixed point \eqref{phi 4 interaction}.
As mentioned previously in this section, perturbative analysis can be performed by assuming $\{\gamma,g_{ij}\}\sim O(\epsilon)$. In the framework of conformal perturbation theory \cite{komargodski2017random}, the one-point function of the singlet operator $\mathcal{O}$ takes the following form:
\begin{equation}
\label{scalar one pt}
\begin{aligned}
\langle\mathcal{O}(y)\rangle=&-\Lambda^{\frac{6\epsilon}{N+8}}\gamma\int d^2z \langle\mathcal{O}(y) \mathcal{O}(z)\rangle + \Lambda^{\frac{12\epsilon}{N+8}}\frac{\gamma^2}{2}\int d^2z_1d^2z_2 \langle\mathcal{O}(y) \mathcal{O}(z_1)\mathcal{O}(z_2)\rangle\\
&+\Lambda^{2\frac{N+6}{N+8}\epsilon}\frac{g_{ij}g_{kl}}{2}\int d^2z_1d^2z_2 \langle\mathcal{O}(y) T^{ij}(z_1)T^{kl}(z_2)\rangle+O\left (\epsilon^3\right)\\
=&-\frac{1}{32 \pi ^3 N |y|^{\Delta_\mathcal{O}}}\left( \gamma -\frac{N+8}{12 \pi  N \epsilon }\gamma^2-\frac{(N+8)}{\pi  (N+3) \epsilon }\text{Tr}\{g^2\}+O\left(\epsilon^2\right)\right).
    \end{aligned}
\end{equation}
Using the minimal subtraction scheme (MS,  see appendix \ref{MS} for details) one obtains the following (minus-)beta function
\begin{equation}
\label{scalar beta}
\begin{aligned}
-\beta(\gamma)=\frac{6\epsilon}{N+8}\gamma-\frac{\gamma^2}{2\pi N}-\frac{2}{\pi}\text{Tr}\{g^2\}+O\left(\epsilon^3\right)\,. 
    \end{aligned}
\end{equation}
Another useful observable is the one-point function of the traceless symmetric operator $T^{ij}$, which reads:
\begin{equation}
\label{stt one pt}
\begin{aligned}
\langle T^{ij}(y)\rangle=&-\Lambda^{\frac{N+6}{N+8}\epsilon}g_{kl}\int d^2z \langle T^{ij}(y) T^{kl}(z)\rangle \\
&+ \Lambda^{\frac{N+12}{N+8}\epsilon}\gamma g_{kl}\int d^2z_1d^2z_2 \langle T^{ij}(y) \mathcal{O}(z_1) T^{kl}(z_2)\rangle\\
&+\Lambda^{2\frac{N+6}{N+8}\epsilon}\frac{g_{kl}g_{mn}}{2}\int d^2z_1d^2z_2 \langle T^{ij}(y) T^{kl}(z_1)T^{mn}(z_2)\rangle+O\left (\epsilon^3\right)\\
=&-\frac{1}{8 \pi ^3 |y|^{\Delta_T}}\left( g_{ij}-\frac{N+8}{6 \pi  N \epsilon } g_{ij} \gamma-\frac{N+8}{\pi (N+6) \epsilon }\left((g^2)_{ij}-\frac{\delta_{ij}}{N}\text{Tr}\{g^2\}\right)\right)\,. 
    \end{aligned}
\end{equation}
Using the MS scheme we find:
\begin{equation}
\label{stt beta}
\begin{aligned}
-\beta(g_{ij})=&\frac{N+6}{N+8}\epsilon g_{ij}-\frac{g_{ij}\gamma}{\pi N}-\frac{1}{\pi  }\left((g^2)_{ij}-\frac{\delta_{ij}}{N}\text{Tr}\{g^2\}\right)+O\left(\epsilon^3\right).
    \end{aligned}
\end{equation}
In the following, we are to investigate the solutions of null beta functions \eqref{scalar beta} and \eqref{stt beta} at the vicinity of the trivial fixed point, such that they are continuously connected to $\gamma=g_{ij}=0$ at the limit $\epsilon \to 0$.

Obviously, for all $N$ there exists a $O(N)$-symmetric fixed point:
\begin{equation}
\label{O(N) FP}
\begin{aligned}
\gamma=\gamma_*\equiv\frac{12\pi N}{N+8}\epsilon+O\left(\epsilon^2\right); \quad g_{ij}=0,
    \end{aligned}
\end{equation}
which is IR-stable against $O(N)$ symmetric deformation. In $O(N)$ preserving cases, while the flow triggered by a $\gamma>0$ deformation ends in the stable fixed point \eqref{O(N) FP}, a $\gamma<0$ one from \eqref{scalar beta} naively implies a flow toward minus infinity. However, of course, this exceeds the perturbative regime and the end point of such a flow is beyond the scope of this section. One possibility for the physical behavior in this regime arises from the EFT description proposed in section \ref{sec_Effective_Action}.

The lowest-lying defect primaries at the fixed point \eqref{O(N) FP} include 
\begin{equation}
\begin{aligned}
\Delta(\hat{\phi}^i)=&1-\frac{N-4  }{2N+16}\epsilon+O\left(\epsilon^2\right),\\
\Delta(\hat{T}^{ij})=&2-\frac{N-6 }{N+8}\epsilon+O\left(\epsilon^2\right).
    \end{aligned}
\end{equation}
The defect order parameters $\{\hat{\phi}^i\}$, not surprisingly, are relevant when $\epsilon\ll1$. However, for $\{T^{ij}\}$ deformations,  the statement will depend on $N$: when $N<6$ the fixed point in \eqref{O(N) FP} is stable, while when $N>6$ it is unstable. When $N=6$, the one-loop analysis result \eqref{stt beta} suggests that the stability depends on the specific deformations that are being triggered. We elaborate on these cases and classify the various deformations in the following subsection \ref{subsec_DCFT_FP_PER}.

Another notable statement is that there exists a single operator, which is $O(N)$-singlet, of $\mathbb{S}^{1-\epsilon}$ spin 1 and protected dimension 3. This is the displacement operator, \footnote{The displacement operator appears in the Ward identities corresponding to the broken translations in the directions orthogonal to the defect, see e.g. \cite{Billo:2016cpy,Cuomo:2021cnb}.}and we will denote it as $\partial_{y}\hat{\mathcal{O}}$ to illustrate that it is continuously connected to the $\mathcal{O}$ level-1 descendant along the RG flow. 

Finally, as a counterpart to the free theory discussion \eqref{eq_defect_central_charge_free}, we present the defect Weyl anomaly. The perturbative result is extensively studied in \cite{shachar2022rg}, and following the results derived therein, together with the fixed point \eqref{O(N) FP}, we find:

\begin{equation}
\begin{aligned}
b_{\text{IR}}=-(2-\Delta_{\mathcal{O}})(\pi \mathcal{N}_\mathcal{O}\gamma_*)^2+O\left(\epsilon^4\right)=-\frac{27 N }{(N+8)^3}\epsilon ^3+O\left(\epsilon^4\right).
    \end{aligned}
\end{equation}

\subsection{Perturbative Fixed Points}
\label{subsec_DCFT_FP_PER}

Following the discussion in the last subsection, we move on to study perturbative fixed points of maximal symmetries that can be preserved by $\{T^{ij}\}$, that is $O(M)\times O(N-M)\subseteq O(N)$. Since beta functions \eqref{scalar beta} and \eqref{stt beta} are quadratic in couplings at the one-loop level, we note this is the only symmetry pattern that can be found from solving null conditions to leading perturbative order. However, in principle, one could calculate higher-loop corrections and obtain a more complicated symmetry group.

We consider the following deformation,  being triggered along with $\mathcal{O}$ as in equation \eqref{eq_action_with_def}:

\begin{equation}
\label{TM def}
\begin{aligned}
g_{ij} T^{ij}=g_M\left[-\frac{1}{M}\sum_{i\leq M}(\phi^i)^2+\frac{1}{N-M}\sum_{i> M}(\phi^i)^2\right],
    \end{aligned}
\end{equation}
for $1\leq M <N$ and with the convention $g_M > 0$.\footnote{A negative $g_{M}$ is equivalent to a positive $g_{N-M}$. This is to avoid over-counting of cases.} The beta function \eqref{stt beta} for $g_M$ reduces to:

\begin{equation}
\label{eq_g_M_beta}
\begin{aligned}
-\beta(g_M)=\frac{N+6}{N+8} \epsilon g_M-\frac{g_M \gamma}{\pi  N}-\frac{2M-N}{\pi M(N-M)}g_M^2+O\left(\epsilon^3\right).
    \end{aligned}
\end{equation}
In what follows, we classify the fixed-point solutions according to cases of $\{M, N\}$. 

When $N<6$, each choice of $M$ admits a single solution to null equations of \eqref{scalar beta} and \eqref{stt beta}. In the space spanned by $\gamma$ and $g_M$, such fixed points have one relevant direction and therefore they are unstable. 

When $N=6$, at the fixed point \eqref{O(N) FP} we have the beta function \eqref{eq_g_M_beta}

\begin{equation}
\begin{aligned}
\left.-\beta(g_M)\right|_{\gamma=\gamma_*,N=6}=\frac{M-3}{M-6}\frac{2 g^2_M}{ \pi M} +O\left(\epsilon^3\right),
    \end{aligned}
\end{equation}
such that for $M>3$ the operator $\hat{T}_M$ is irrelevant, and one meta-stable fixed point can be found, similarly to the cases of $N<6$. For $M=3$, the meta-stable fixed point collides with the $O(N)$-symmetric one of equation \eqref{O(N) FP} at the one-loop level, and the solution is sensitive to higher-loop corrections. For $M<3$, on the other hand, null equations of \eqref{scalar beta} and \eqref{stt beta} yield no finite positive solution to $g_M$. We note in these cases $\hat{T}_M$ at the fixed-point \eqref{O(N) FP} is relevant, and the RG being triggered flows toward large couplings, where the perturbative scheme breaks down. We will therefore categorize $\{M>3, N=6\}$ cases as in the same class of $N<6$, where the single $O(M)\times O(N-M)$ fixed point is unstable to one deformation, and present the data in table \ref{N<=6 fixed point}.

\begin{table}[!h]
\begin{center}
\begin{tabular}{ |c|c|c|c|c|c| } 
 \hline
$((g_{M})_*,\gamma_{*})$ & $N=2$ & $N=3$ & $N=4$ & $N=5$ & $N=6$\\ 

  \hline
$M=1$ & $\left(\frac{\sqrt{2}}{5},\frac{8}{5}\right)$ & $\left( \frac{4 \sqrt{7}-2}{33},\frac{26+2 \sqrt{7}}{11} \right)$ & $\left( \frac{\sqrt{6}-1}{8}, \frac{9+\sqrt{6}}{3}\right)$ & $\left(\frac{8 \sqrt{5}-12}{65},\frac{46 + 6 \sqrt{5}}{13} \right) $ & No FP \\ 

  \hline
$ M=2$ &  & $\left(\frac{4 \sqrt{7}+2}{33},\frac{26-2 \sqrt{7}}{11} \right)$ & $\left(\frac{\sqrt{5}}{6},\frac{10}{3}\right)$ & $\left(\frac{12 \sqrt{3}-6}{65},\frac{54 + 2 \sqrt{3}}{13} \right)$ & No FP\\ 
\hline

$M=3$ &  &  & $\left( \frac{\sqrt{6}+1}{8}, \frac{9-\sqrt{6}}{3}\right)$ & $\left(\frac{12 \sqrt{3}+6}{65},\frac{54 - 2 \sqrt{3}}{13} \right)$ & ? \\ 
\hline
$M=4$ &  &  & & $\left(\frac{8 \sqrt{5}+12}{65},\frac{46 - 6 \sqrt{5}}{13} \right)$ & $\left(\frac{8}{21},\frac{32}{7} \right)$  \\ 
\hline
$M=5$ &  &  & &  & $\left(\frac{10}{21},\frac{20}{7} \right)$ \\ 
\hline
\end{tabular}
\caption{\label{N<=6 fixed point} Fixed points for $N\leq 6$ (in addition to the trivial and $O(N)$ symmetric ones).The couplings are in units of $\pi \epsilon$ and only the leading orders are presented. At $N=6$ and $M=3$, the fixed point existence needs to be discussed at a higher-loop level.}
\end{center}
\end{table}

When $N>6$, most choices of $\{M, N\}$ admit no perturbative fixed point other than the trivial one and the $O(N)$-symmetric one \eqref{O(N) FP}, similar to $\{M<3, N=6\}$. Exceptions exist for $\{M=N-1, 6<N<10\}$, where there are two $O(N-1)\times \mathbb{Z}_2$ fixed points. We will denote the unstable one as FP-1 and the stable one as FP-2. When $N=10$, FP-1 and FP-2 collide at the 1-loop level, and higher loop corrections are needed to specify the existence and number of the fixed point. Data corresponding to the aforementioned cases are presented in table \ref{N>6 fixed point}.

\begin{table}[!h]
\begin{center}
\begin{tabular}{ |c|c|c|c|c|c| } 
 \hline
$((g_{M})_*,\gamma_{*})$ & $N=7$ & $N=8$ & $N=9$ & $N=10$ \\ 

  \hline
FP-1 & $\left(\frac{10+4 \sqrt{3}}{35}, \frac{66-10 \sqrt{3}}{15} \right)$ & $\left(\frac{21+7 \sqrt{2}}{64}, \frac{19-3 \sqrt{2}}{4} \right)$ & $\left( \frac{8}{17}, \frac{72}{17}\right)$ & ?  \\ 

  \hline
FP-2 & $\left(\frac{10-4 \sqrt{3}}{35}, \frac{66+10 \sqrt{3}}{15} \right)$ & $\left(\frac{21-7 \sqrt{2}}{64}, \frac{19+3 \sqrt{2}}{4} \right)$ & $\left( \frac{40}{153}, \frac{100}{17}\right)$ & ? \\ 
\hline
\end{tabular}
\caption{\label{N>6 fixed point}Fixed points for $\{M=N-1, 6<N\leq 10\}$ (in addition to the trivial and $O(N)$ symmetric ones). The couplings are in units of $\pi \epsilon$ and only the leading orders are presented. At  $\{M=9, N=10\}$, FP-1 and FP-2 collide at the 1-loop level and higher loops are needed to distinguish them.}
\end{center}
\end{table}

\begin{figure}[!h]
\centering
  \includegraphics[width=0.32\textwidth ]{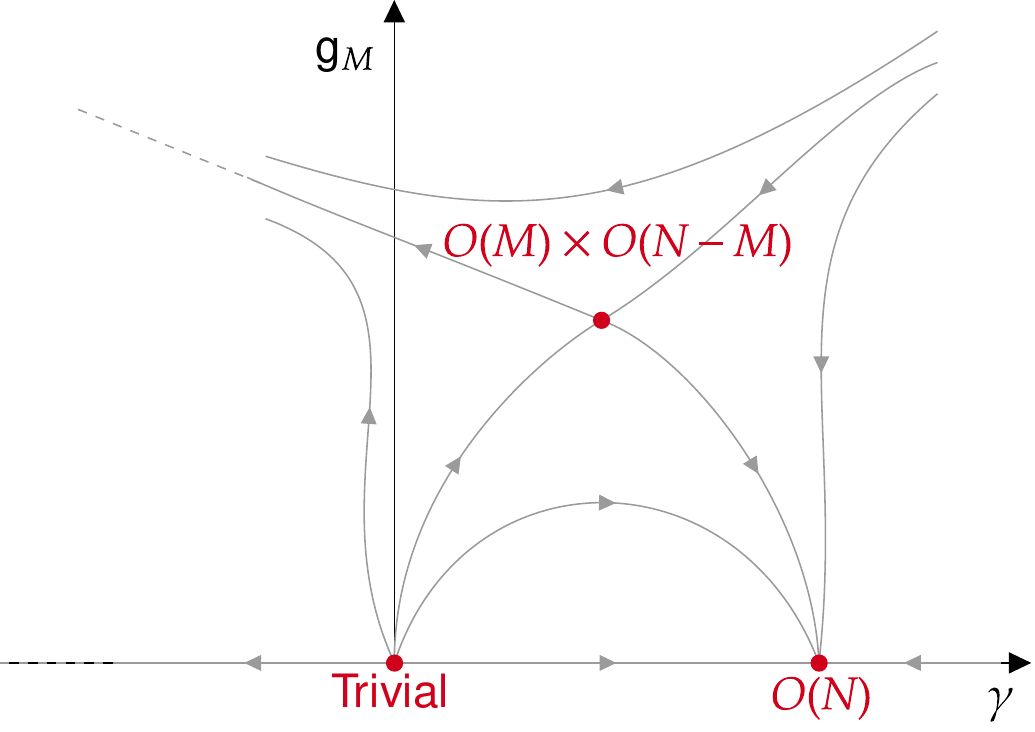}
    \includegraphics[width=0.32\textwidth ]{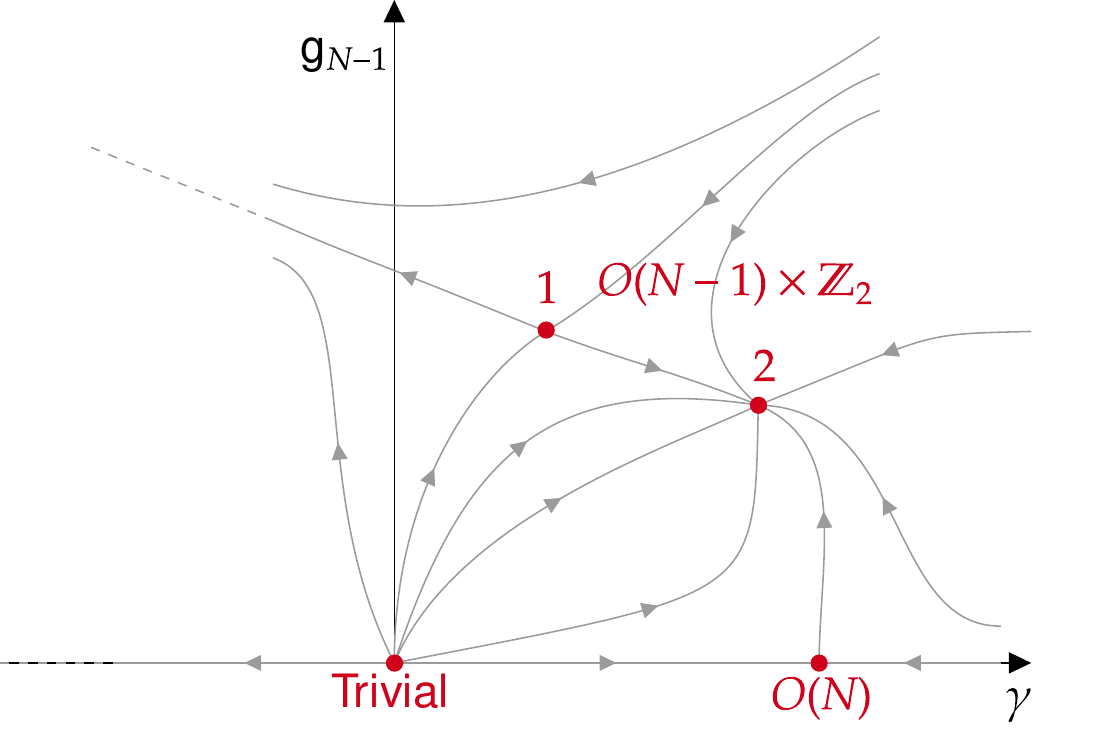}
      \includegraphics[width=0.32\textwidth ]{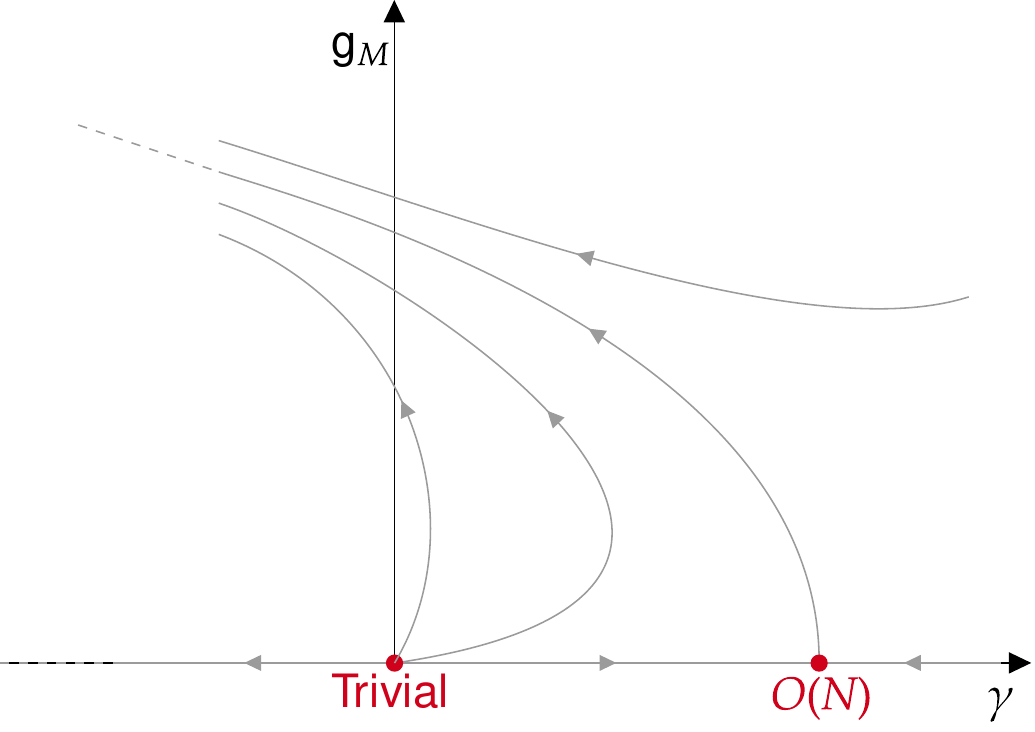}
  \caption{\label{Phase Diagram}The perturbative phase diagram for $O(M)\times O(N-M)$ defects. Left: cases listed in table \ref{N<=6 fixed point}. Middle: cases listed in table \ref{N>6 fixed point}. Right: cases of no fixed point other than the trivial one and $O(N)$-symmetric one. }
\end{figure}

To conclude this part, we have obtained the perturbative phase diagram (see figure \ref{Phase Diagram} for an illustration of cases) and the one-loop perturbative fixed points within. It is a physically interesting question to ask what is the IR theory at the end of the asymptotic directions described above. This will be further studied in the next section.

\section{Effective Action and Phase Diagram}
\label{sec_Effective_Action}
In the following, we discuss possible effective actions of the non-perturbative fixed points mentioned in the last section and propose completion of the phase diagram from figure \ref{Phase Diagram}. Classically, since $g_M$ and $-\gamma$ flow to large values as suggested by \eqref{scalar beta} and \eqref{stt beta}, the defect has a tendency to acquire localized degrees of freedom. Inspired by similar studies \cite{krishnan2023plane,metlitski2022boundary}, we make the assumption that the effective actions consist of three parts:

\begin{itemize}
  \item A DCFT of the background theory $S_{\mathcal{D}}$.
  \item An action associated with the localized defect degrees of freedom $S_{\sigma}$.
  \item Couplings between the two theories $S_{\mathcal{D}}$ and $S_{\sigma}$.
\end{itemize}
We will further assume that $S_{\sigma}$ can be described by the two-dimensional Non-Linear Sigma Model (NL$\Sigma$M) \cite{zinn2021quantum,henriksson2023critical}. This follows from the consideration that symmetry patterns as discussed in the previous section are $O(M)\times O(N-M)$, among which classically a $O(M)$ order parameter is favored by a large coupling $g_M$. 
\footnote{For a recent study on the subject of spontaneous symmetry breaking on surface defects, see \cite{2023Cuomo}.}
Consistency check and verification of such an effective action proposal will be the main subject of this section. 

For simplicity, we will use a coordinate system slightly different from previous ones throughout this section. Let $y\in \mathbb{R}^+$ be the distance to the defect and $\Omega^{d-3}\in \mathbb{S}^{d-3}$ be the compact notation of the spherical coordinates, such that $x=(y, \Omega^{d-3},z)$. We remind readers that the background theory is the same as in section \ref{sec_TheON_critical_perturbationTheory}: the $O(N)$ Wilson-Fisher fixed point at $\epsilon \equiv 4-d \ll 1$. The fact that the background theory, contrary to the defect theory, is perturbative will be essential in our following analysis. 

This section is organized as follows. In subsection \ref{subsec_Mean_Field}, we identify the saddle point of $S_{\mathcal{D}}$ and perform a perturbative analysis around it. In subsection \ref{subsec_Tilt_Data}, we extract DFCT data of our interest from $S_{\mathcal{D}}$. Finally, in subsection \ref{subsec_NLSM}, we show that the coupling between $S_{\mathcal{D}}$ and $S_\sigma$ is unique and verify the NL$\Sigma$M stability.

\subsection{Mean Field and $\mathbb{H}^3\times \mathbb{S}^{d-3}$}
\label{subsec_Mean_Field}

We will take $S_{\mathcal{D}}$ to be the surface analog of the line defect induced by a localized magnetic field in $O(N)$ Wilson-Fisher fixed point \cite{Cuomo:2021rkm}, where the background order parameter $\phi^i$ acquires a non-zero vacuum expectation value when being evaluated close to the defect. We note that similar problems were also studied in the context of the boundary universality class \cite{burkhardt1987surface,mcavity1995conformal,shpot2021boundary,dey2020operator, Padayasi:2021sik}. For our purpose, we will study the perturbative expansion around the mean-field profile. The profile is subject to the classical equation of motion, which reads:
\begin{equation}
\label{mean field eom}
\begin{aligned}
\left(-\partial^2_{\mathbb{R}^d}+\frac{\lambda_\text{b}}{6}\left(\phi_{\text{cl}}(x)\right)^2\right)\phi_{\text{cl}}(x)=0\,.
    \end{aligned}
\end{equation}
In addition to the trivial solution $\phi_{\text{cl}}(x)=0$, equation \eqref{mean field eom} admits a profile that is singular when evaluated close to the defect at $y=0$: 
\begin{equation}
\label{eq_classical profile}
\begin{aligned}
\phi_{\text{cl}}(x)=\sqrt{\frac{6(1+\epsilon)}{\lambda_\text{b}}}\frac{1}{y}\,.
    \end{aligned}
\end{equation}
We then facilitate a field redefinition and study the fluctuations on top of the configuration \eqref{eq_classical profile}. Explicitly, we define the fluctuation fields $\{\Tilde{\phi}^i\}$ according to:
\begin{equation}
\begin{aligned}\label{eq_O(N) breaking}
\phi^i(x)=\delta^{i1}\phi_{\text{cl}}(x)+\Tilde{\phi}^i(x).
    \end{aligned}
\end{equation}
The action of the background fields \eqref{bulk model} can be recasted into the classical part and a functional of the fluctuation fields, which reads:
\begin{equation}
\label{eq_ mean field action}
    \begin{aligned}
       S_0=&\int_{y>\frac{1}{\Lambda}}d^d x \left\{\frac{1}{2}(\partial \phi^i)^2+\frac{\lambda_\text{b}}{4!}[(\phi^i)^2]^2\right\}\\
       =&S_{\text{cl}}+\int_{y>\frac{1}{\Lambda}}d^d x \left\{\frac{1}{2}(\partial \Tilde{\phi}^i)^2+\frac{1+\epsilon}{2y^2}\left[(\Tilde{\phi}^i)^2+2(\Tilde{\phi}^1)^2\right]\right.\\
       &\left.+\sqrt{\frac{(1+\epsilon)\lambda_\text{b}}{6}}\frac{\Tilde{\phi}^1}{y }(\Tilde{\phi}^i)^2+\frac{\lambda_\text{b}}{4!}\left[(\Tilde{\phi}^i)^2\right]^2\right\}.
    \end{aligned}
\end{equation}
In order to perform perturbative analysis with respect to the coupling $\lambda_{\text{b}}$, it is useful to compute the saddle point propagator. We digress here to discuss generally how this can be obtained. Consider the Green's function $G(x,{x}')$ subject to the following equation:
\begin{equation}
\label{eq_saddle point eom}
\begin{aligned}
\left(-\partial^2_{\mathbb{R}^d}+\frac{(6-d)(d-2)+4m^2}{4 y^2}\right)G(x,{x}')=\delta_{\mathbb{R}^d}(x-{x}').\, 
    \end{aligned}
\end{equation}
with $m^2\in \mathbb{R}$ being a constant. This problem can be effectively solved by a Weyl transformation from $\mathbb{R}^d$ to $\mathbb{H}^3\times\mathbb{S}^{d-3}$ (see e.g. \cite{nishioka2021free,Cuomo:2021rkm}): 
\begin{equation}
\begin{aligned}
d^2 s=&d^2z+d^2 y+y^2d^2\Omega^{d-3}\\
=&y^2\left(\frac{d^2z+d^2 y}{y^2}+d^2\Omega^{d-3}\right)\\
=&y^2 d^2\Tilde{s} \rightarrow d^2\Tilde{s}\,.
    \end{aligned}
\end{equation}
We will denote the Weyl transformation of $G(x,{x}')$ as $\tilde{G}(x,{x}')$. One finds: 
\begin{equation}
\begin{aligned}
(-\partial^2_{\mathbb{H}^3}-\partial^2_{\mathbb{S}^{1-\epsilon}}+m^2)\Tilde{G}(x,{x}')=\delta_{\mathbb{H}^3\times \mathbb{S}^{1-\epsilon}}(x-{x}').
    \end{aligned}
\end{equation}
where $m^2$ has a physical interpretation of mass term in $\mathbb{H}^3\times\mathbb{S}^{d-3}$. To solve the above, we can first perform an expansion by $\mathbb{S}^{1-\epsilon}$ spherical harmonics $Y_{l,q}(\Omega)$:
\begin{equation}
\begin{aligned}
\Tilde{G}(x,{x}')=\sum_{l\geq 0,q}Y_{l,q}(\Omega)Y^*_{l,q}({\Omega}') \Tilde{G}(\xi ;l)\,,
    \end{aligned}
\end{equation}
where
\begin{equation}
\begin{aligned}
-\partial^2_{\mathbb{S}^{1-\epsilon}}Y_{l,q}(\Omega)=&l(l-\epsilon)Y_{l,q}(\Omega),\\
\sum_{q} 1=&\text{deg}(1-\epsilon,l),
    \end{aligned}
\end{equation}
and the degeneracy is given by equation \eqref{eq_harmonics_degeneracy}. In the above,  the dependence on the variable $\xi$ is fixed by the $\mathbb{H}^3$ isometry:
\begin{equation}
\begin{aligned}
\xi=\frac{2 y {y}'}{y^2+{y}'^2+(z-{z}')^2}.
    \end{aligned}
\end{equation}
The solution for each $\mathbb{S}^{1-\epsilon}$ spin $l$ functions $\Tilde{G}(\xi ;l)$ is given by the standard Euclidean $AdS_3$ bulk to bulk propagator, which has been well studied \cite{d2004supersymmetric}, and reads:
\begin{equation}
\begin{aligned}
\Tilde{G}(\xi ;l)=&\frac{ 1}{2\pi
}\left(\frac{\xi}{2}\right)^{\Delta_l}\, _2F_1\left(\frac{\Delta_l }{2},\frac{\Delta_l +1}{2};\Delta_l ;\xi ^2\right),
    \end{aligned}
\end{equation}
where $\Delta_l$ are given by $\Delta_l\equiv 1+\sqrt{1+m^2+l(l-\epsilon)}$. These correspond to the (non-singular at the $\xi\to 0$ limit) solution to the conformal boundary condition, as follows from the analysis in \cite{nishioka2021free}.

Notice that since the defect breaks the Poincare symmetry and introduces a scale $y$, the Green function in this case acquires a non-zero value at the coincident point after regulation. Let $\text{vol}(\mathbb{S}^{1-\epsilon})$ be the volume of the unit sphere $\mathbb{S}^{1-\epsilon}$, one finds: 
\begin{equation}
\begin{aligned}
\Tilde{G}(x,x)=&\sum_{l\geq 0}\frac{\deg (1-\epsilon,l)}{ \text{vol}(\mathbb{S}^{1-\epsilon})}\lim_{\xi \rightarrow 1}\Tilde{G}(\xi ;l)\\
=&\sum_{l\geq 0}\frac{\deg (1-\epsilon,l)}{\text{vol}(\mathbb{S}^{1-\epsilon})}\lim_{\xi \rightarrow 1}\left(\frac{i}{4 \sqrt{2} \pi  \sqrt{\xi -1}}+\frac{1-\Delta_l}{4 \pi
   }+O\left(\sqrt{\xi -1}\right)\right)\\
=&-\sum_{l\geq 0}\frac{\deg (1-\epsilon,l)}{4\pi \text{vol}(\mathbb{S}^{1-\epsilon})}\sqrt{1+m^2+l(l-\epsilon)},\\
    \end{aligned}
\end{equation}
where in the last equation, we have used the fact that $\sum_{l\geq 0}\deg (1-\epsilon,l)=0$ under dimensional regulation. To solve the above expression to its leading order, we apply a method similar to that in \cite{cuomo2022localized} and expand summation terms around $l=\infty$
\begin{equation}
\label{eq_reg}
\begin{aligned}
\Tilde{G}(x,x)=&-\frac{\pi ^{\frac{\epsilon }{2}} \Gamma
   \left(1-\frac{\epsilon }{2}\right)}{8\pi^2} +\sum_{l\geq 1}\left(-\frac{2^{\epsilon} \pi ^{\frac{\epsilon -1}{2}}l^{1-\epsilon}}{4\pi \Gamma \left(\frac{1}{2}-\frac{\epsilon }{2}\right)}-\frac{2^{\epsilon -3} \pi ^{\frac{\epsilon -3}{2}} (\epsilon -1) \epsilon l^{-\epsilon}}{\Gamma \left(\frac{1}{2}-\frac{\epsilon }{2}\right)}\right.\\
   &\left.-\frac{2^{\epsilon -5} \pi ^{\frac{\epsilon -3}{2}} \left(3 \epsilon ^4-2 \epsilon ^3+2 \epsilon +12+12m^2\right)l^{-1-\epsilon}}{3  \Gamma
   \left(\frac{1}{2}-\frac{\epsilon }{2}\right)}+O\left(l^{-2-\epsilon}\right)\right)\\
   =&-\frac{2^{\epsilon -5} \pi ^{\frac{\epsilon -3}{2}} \left(3 \epsilon ^4-2 \epsilon ^3+2 \epsilon +12+12m^2\right)}{3
   \Gamma \left(\frac{1}{2}-\frac{\epsilon }{2}\right)}\zeta (\epsilon +1)+O\left(\epsilon^0\right)\\
   =&-\frac{1+m^2}{8\pi^2 \epsilon}+O\left(\epsilon^0\right)\,. 
    \end{aligned}
\end{equation}
In the above, between the first and second equations we have used dimensional regularization to replace the summation in $l$ with a summation over Riemann zeta functions. Going from the second to the third equations, we notice that the zeta function has a single simple pole at $1$. Under an expansion in $\epsilon\ll 1$, the leading contribution will be from the $\zeta(\epsilon+1)$ term in the summation.

\begin{figure}[!h]
\centering
  \includegraphics[width=0.85\textwidth ]{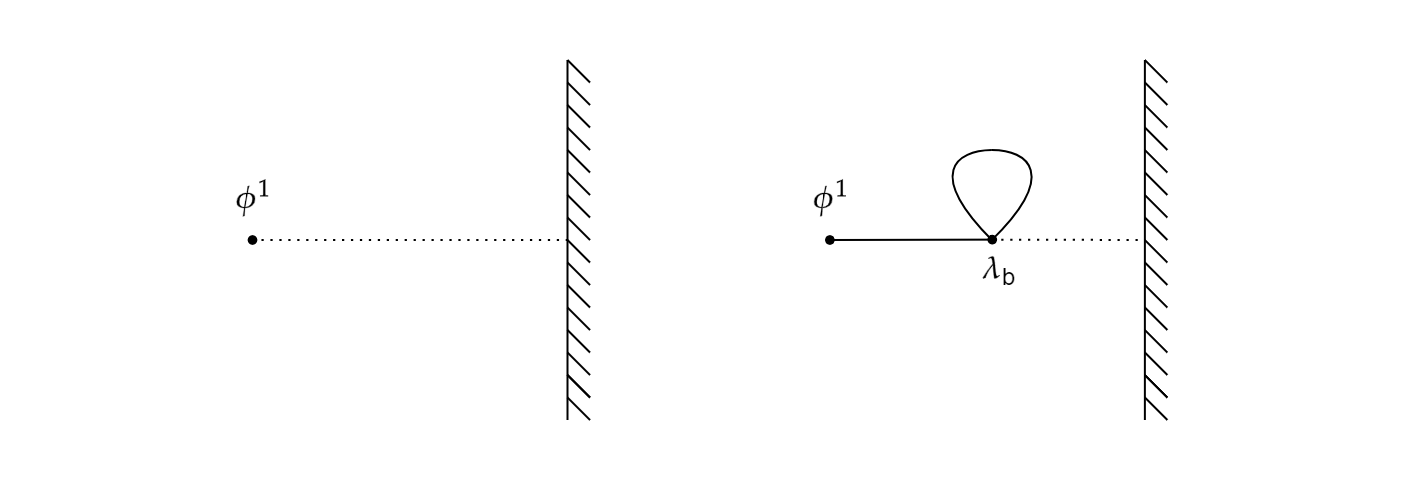}
  \caption{One-loop Feynman diagrams that contribute to equation \eqref{eq_bare calculation}. Dashed lines represent the mean-field profile \eqref{eq_classical profile}, and solid lines represent the Green's function $G_{\text{L}/\text{T}}(x,{x}')$.\label{fig_feynman}}
\end{figure}

Finally, we perform the inverse Weyl transformation to obtain Green's function of the original theory in flat space with a plane defect, which reads:
\begin{equation} \label{Greens function}
\begin{aligned}
G(x,{x}')=&\frac{\Tilde{G}(x,{x}')}{(y{y}')^{d/2-1}}.\\
    \end{aligned}
\end{equation}
Comparing equation \eqref{eq_saddle point eom} with the mean-field saddle point action \eqref{eq_ mean field action}, for the longitudinal (L) mode $\Tilde{\phi}^1$ and transversal (T) modes $\Tilde{\phi}^{i\geq 2}$ we correspondingly find:
\begin{equation}
\begin{aligned}
m_\text{L}^2=&2+3 \epsilon+\frac{\epsilon ^2}{4},\\
m_\text{T}^2=&\epsilon+\frac{\epsilon ^2}{4}.
    \end{aligned}
\end{equation}
The associated Green's function will be denoted by $G_{\text{L}/\text{T}}(x,{x}')$ respectively.
The DCFT $S_\mathcal{D}$ of interest has the following one-point expectation value:
\begin{equation}
\label{eq_expqct}
\begin{aligned}
\langle\phi^i(x)\rangle=&\delta^{i1}\frac{a_\sigma}{y^{\Delta_\phi}}\,  ,\\
    \end{aligned}
\end{equation}
where the scaling dimension is given by  \cite{kehrein1995spectrum,henriksson2023critical}:
\begin{equation}
\begin{aligned}
\Delta_\phi=&1-\frac{\epsilon}{2}+O\left(\epsilon^2\right).
    \end{aligned}
\end{equation}
In the following, we illustrate how this is manifested by the mean-field analysis to the 1-loop level. Consider the formal expansion of $\langle\phi^1(x)\rangle$ as (see figure \ref{fig_feynman}):
\begin{equation}
\label{eq_bare calculation}
    \begin{aligned}
        \langle\phi^1(x)\rangle=&\phi_{\text{cl}}(x)-\frac{\lambda_\text{b}}{2}\int d^d {x}'G_\text{L}(x,{x}')G_\text{L}({x}',{x}')\phi_{\text{cl}}({x}')\\
        &-\frac{N-1}{6}\lambda_\text{b}\int d^d {x}'G_\text{L}(x,{x}')G_\text{T}({x}',{x}')\phi_{\text{cl}}({x}')+O\left(\lambda_\text{b}^{3/2}\right)\\
        =&\sqrt{\frac{6(1+\epsilon)}{\lambda_\text{b}}}\frac{1}{y}+\sqrt{\frac{(1+\epsilon)\lambda_\text{b}}{6}}\frac{1}{y^{1-\epsilon}}\left [\frac{N+8}{16 \pi ^2 \epsilon} +O\left(\epsilon^0\right)\right]+O\left(\lambda_\text{b}^{3/2}\right),\\
    \end{aligned}
\end{equation}
where the bare coupling $\lambda_\text{b}$ in terms of the cut-off $\Lambda$ and renormalzied coupling $\lambda$ reads:
\begin{equation}
\label{fp2}
\begin{aligned}
\lambda_\text{b}=\Lambda^{\epsilon}\left(\lambda+\frac{N+8}{48\pi^2\epsilon}\lambda^2+O\left(\lambda^3\right)\right).
    \end{aligned}
\end{equation}

From the above we find that indeed the pole term in \eqref{eq_bare calculation} cancels, in agreement with the regulation scheme in \eqref{eq_reg} and the perturbative analysis: 
\begin{equation}
    \begin{aligned}
        \langle\phi^1(x)\rangle =\sqrt{\frac{6(1+\epsilon)}{\lambda}}\frac{1}{\Lambda^{\epsilon/2}y}+\sqrt{\frac{\lambda}{6}}\left[\frac{N+8}{16\pi^2}\log \left(y \Lambda\right)+O\left(\epsilon\right)\right]+O\left(\lambda^{3/2}\right).\\
    \end{aligned}
\end{equation}
Finally, plugging the fixed point value \cite{kehrein1995spectrum,henriksson2023critical}:
\begin{equation}
    \begin{aligned}
        \lambda=\lambda_*= \frac{48\pi^2\epsilon}{N+8}\left(1+\frac{3 (3 N+14) }{(N+8)^2} \epsilon+O\left(\epsilon ^2\right)\right)\, ,
    \end{aligned}
\end{equation}
We find $\langle\phi^1(x)\rangle$ to the leading order in $\epsilon$ has the expected scaling dimension as in equation \eqref{eq_expqct}, and $a_\sigma$ reads:
\begin{equation}
    \begin{aligned}\label{asigma}
        a_\sigma=\frac{1}{2\pi}\sqrt{\frac{N+8}{2\epsilon}}\left[1+\frac{N^2+7N+22}{2(N+8)^2}\epsilon+O\left(\epsilon^2\right)\right].
    \end{aligned}
\end{equation}

\subsection{DCFT Data}
\label{subsec_Tilt_Data}
As locality requires, the assumed couplings between $S_{\sigma}$ and $S_{\mathcal{D}}$ should be between defect degrees of freedom and those of the neighboring background layer. We will discuss $S_{\sigma}$ in the next subsection, and elaborate on the layer of background degrees of freedom, which is described by the DCFT data, in what follows. Such data can be approached via the bulk-to-defect OPE \cite{liendo2013bootstrap, gaiotto2014bootstrapping}, and for the background order parameters $\{\phi^i\}$ it reads: 
\begin{equation}
\label{bulk to defect OPE}
\begin{aligned}
\phi(x)=\sum_{\hat{\mathcal{O}}}C_{\hat{\mathcal{O}}}^{\phi}\frac{Y^*_{s_{\hat{\mathcal{O}}}}(\Omega)}{y^{\Delta_\phi-\Delta_{\hat{\mathcal{O}}}}}\mathcal{B}_{\Delta_{\hat{\mathcal{O}}}}(y,\partial)\hat{\mathcal{O}}(z),
    \end{aligned}
\end{equation}
where  $\{\hat{\mathcal{O}}\}$ are defect primaries of $\mathbb{S}^{1-\epsilon}$ spin $s_{\hat{\mathcal{O}}}$ and scaling dimension $\Delta_{\hat{\mathcal{O}}}$. Spin- and $O(N)$-indices have been omitted for simplicity and will be restored when necessary. $\mathcal{B}_{\Delta}$ is fixed by the conformal symmetry (or equivalently, the $\mathbb{H}^3$ isometry) up to multiplication factors, which are denoted as OPE coefficients $\{C_{\hat{\mathcal{O}}}^{\phi}\}$. One finds \cite{gaiotto2014bootstrapping}:

\begin{equation}
\begin{aligned}
\mathcal{B}_{\Delta}(y,\partial)=\sum_{n\geq 0}\frac{(-1)^ny^{2n}}{n!(\Delta)_{n}}(\partial^2_{z^1}+\partial^2_{z^2})^n, 
    \end{aligned}
\end{equation}
where $(\Delta)_n$ is the Pochhammer symbol. In the representation of the defect conformal group $O(3,1)$, $\{\phi^i\}$ are of spin-zero, and therefore only scalar primaries will be included on the RHS of \eqref{bulk to defect OPE}. We choose the following normalization convention for the two-point functions of scalar defect primaries:
\begin{equation}
\label{defect normalization}
\begin{aligned}
\langle\mathcal{\hat{O}}_a (z_1)\mathcal{\hat{O}}_b(z_2)\rangle=&\frac{\delta_{ab}}{|z_1-z_2|^{2\Delta_{\mathcal{\hat{O}}_a}}}.
    \end{aligned}
\end{equation}
Next, we compute the two-point functions $\langle \phi^i(x)\phi^i({x}')\rangle$ (considered without a summation over the repeated index $i$). We use the result from subsection \ref{subsec_Tilt_Data} together with the decomposition in equation \eqref{bulk to defect OPE} to obtain the DCFT data. Generally, the Green's function \eqref{Greens function} in the limit $\xi\to 0$ reads: 

\begin{equation}
\begin{aligned}
G(x,{x}')=&\frac{1}{(y {y}')^{d/2-1}}\sum_{l\geq 0,q}Y_{l,q}(\Omega)Y^*_{l,q}({\Omega}')\left(\frac{\xi}{2}\right)^{\Delta_l}\left(\frac{1}{2\pi}+O\left(\xi^2\right)\right). \\
    \end{aligned}
\end{equation}
For $i \geq 2$, the leading order
$\langle \phi^i(x)\phi^i({x}')\rangle$ is given by Green’s function $G_{\text{T}}(x,{x}')$. From the above, we can conclude that for $s_{\hat{\mathcal{O}}}=l$: 
\begin{align}
\Delta_{\hat{\mathcal{O}}}=&1+\sqrt{l^2+1}+O\left(\epsilon \right),\label{eq_dcft_data}\\
C_{\hat{\mathcal{O}}}^{\phi}=&\frac{1}{\sqrt{2 \pi}}+O\left(\epsilon \right).\label{DCFT data}
    \end{align}
Therefore, in the decomposition \eqref{bulk to defect OPE} for $\phi^i$ with $i\geq2$, defect primaries to their leading order $O(\epsilon^0)$ are of dimension $\Delta_{\hat{\mathcal{O}}}\geq 2 $. We note that the lowest-lying operator has a protected dimension: Since the saddle point \eqref{eq_O(N) breaking} breaks the $O(N)$-symmetry, it is to be identified with the `tilt' operator $Q$ (see e.g. \cite{krishnan2023plane,cuomo2022localized,gaiotto2014bootstrapping}).
Restoring the $O(N)$-indices, such operators are defined (up to normalization) with respect to the background symmetry current $J_\mu^{[ij]}$: 
\begin{equation}
\begin{aligned}
\partial^\mu J_\mu^{[1i]}(x)\propto Q^i(z)\delta^{d-2}(y,\Omega). 
    \end{aligned}
\end{equation}
In the following, we will omit the index $i$ and denote by $\langle Q(z)Q({z}')\rangle$ the two-point function of tilt operators, which takes the form of equation \eqref{defect normalization}. We will also denote by $b_t$ the $\phi$-to-$Q$ OPE coefficient as in equation \eqref{DCFT data}.

The two-point function $\langle\phi^1(x)\phi^1({x}')\rangle$, on the other hand, has a disconnected contribution in addition to Green's function $G_{\text{L}}(x,x')$. The lowest-lying operator in the decomposition of $\phi^1$ is the identity, with the OPE coefficient being $a_\sigma$ in equation \eqref{asigma}. Other defect primaries for $s_{\hat{\mathcal{O}}}=l$ include:

\begin{align}
\Delta_{\hat{\mathcal{O}}}=&1+\sqrt{l^2+3}+O\left(\epsilon \right),\label{eq_dcft_data}\\
C_{\hat{\mathcal{O}}}^{\phi^1}=&\frac{1}{\sqrt{2 \pi}}+O\left(\epsilon \right).\label{DCFT data}
    \end{align}
We note that to the leading order $O(\epsilon^0)$, the $S^{1-\epsilon}$ spin 1 operator in this case has $\Delta_{\hat{\mathcal{O}}}=3$. Similarly to the perturbative case in section \ref{sec_TheON_critical_perturbationTheory}, this defect operator is identified as the displacement operator and has a protected dimension \cite{Billo:2016cpy, Cuomo:2021cnb}.

To conclude this subsection, the DCFT data of $S_{\mathcal{D}}$ that will be needed in what follows are given by:
\begin{equation}
\label{phi OPE}
\begin{aligned}
\phi^1(x)=&\frac{a_\sigma}{y^{\Delta_\phi}}+\cdots\\
\phi^i(x)=&\frac{b_t}{y^{\Delta_\phi-2}}Q^i(z)+\cdots \, ,\qquad \text{for } i\geq 2,
    \end{aligned}
\end{equation}
where the dots stand for defect descendants and primaries of higher dimensions.

\subsection{Coupling to NL$\Sigma$M}
\label{subsec_NLSM}
In this subsection, we turn to discuss defect degrees of freedom which, as part of our assumption, are described by an $O(M)$-NL$\Sigma$M.  
The action $S_\sigma$ for the localized defect degrees of freedom is taken to be:

\begin{equation}
\label{NLSM action}
\begin{aligned}
S_{\sigma }=\frac{1}{2g}\int_{\mathbb{R}^2}d^2z(\partial \Vec{n})^2,
    \end{aligned}
\end{equation}
where $\Vec{n}\in \mathbb{S}^M$ and $g\geq 0$ is the NL$\Sigma$M coupling. We will additionally assume that $g\ll 1$ to generate the perturbative analysis, under which its RG flow can be investigated. 
In what follows, we will introduce the coupling between $S_\sigma$ and $S_{\mathcal{D}}$. We will then verify that $g\ll 1$ is indeed stable at the IR and that the effective action proposal is self-consistent. The method being applied in this subsection mainly follows the those developed in \cite{metlitski2022boundary, Cuomo:2022xgw} \footnote{Note that in  \cite{Cuomo:2022xgw} the theory flows to a line DCFT in the IR, here as we will soon claim the theory flows to the extraordinary-log phase universality class, in analogy with \cite{metlitski2022boundary}.}. 

The NL$\Sigma$M degrees of freedom can be rephrased in terms of $M-1$ scalar fields $\{\pi^i\}$, and we use the following convention: 
\begin{equation}
\begin{aligned}
(\Vec{n})^i=&\pi^i, \qquad \text{for } 2\leq i \leq M,\\
(\Vec{n})^1=&\sqrt{1-(\vec\pi)^2},
    \end{aligned}
\end{equation}
where $(\vec{\pi})^2\equiv \pi^i\pi_i$. We learned from equations \eqref{DCFT data} and \eqref{phi OPE} that the lowest-lying operators in the decomposition of background order parameters are the identity and the tilt operators. Therefore,  there exists a single classically marginal coupling between $S_\sigma$ and $S_{\mathcal{D}}$. The proposed IR-effective action reads:
\begin{equation}
\label{IR effective}
\begin{aligned}
S_{\text{IR}}=S_{\mathcal{D}}+S_{\sigma}-\Tilde{\gamma}\int_{\mathbb{R}^2}d^2 z \sum_{i\geq 2}^M \pi^i Q^i,
    \end{aligned}
\end{equation}
where $\tilde{\gamma}$ is the effective coupling. As in \cite{metlitski2022boundary}, the coupling $\Tilde{\gamma}$ is fixed by requiring the $O(M)$ symmetry not explicitly broken.
Under the assumption $g\ll 1$, where the NL$\Sigma$M fluctuations are suppressed, we could see this by investigating the rotation
\begin{equation}
\begin{aligned}
\left\{\begin{array}{lr}
 \langle\phi^{i=1}(x)\rangle=a_\sigma/y^{\Delta_\phi}\\
 \langle\phi^{i=2}(x)\rangle=0\\
 \pi^{i=2}=0\\
\end{array}\right.\xrightarrow[\text{ rotation}-\theta]{O(M)}\left\{\begin{array}{lr}
 \langle\phi^{i=1}(x)\rangle=a_\sigma\cos{\theta}/y^{\Delta_\phi}\\
 \langle\phi^{i=2}(x)\rangle=a_\sigma\sin{\theta}/y^{\Delta_\phi}\\
 \pi^{i=2}=\sin{\theta}\\\end{array}\right.
    \end{aligned}
\end{equation}
Matching the expectation value of $\phi^{i=2}$ before and after the rotation, one obtains (note that we work in the conventions in which the tilt operator is normalized according to \eqref{defect normalization}):
\begin{equation}
\begin{aligned}
\frac{a_\sigma}{y^{\Delta_\phi}}=&\Tilde{\gamma}\int_{\mathbb{R}^2}d^2{z}'\langle\phi^{i=2}(x)Q^{i=2}({z}')\rangle\\
=&\frac{\Tilde{\gamma}b_t}{y^{\Delta_\phi}}\int_{\mathbb{R}^2}d^2z\left(\frac{y}{y^2+z^2}\right)^2=\frac{\pi \Tilde{\gamma}b_t}{y^{\Delta_\phi}},
    \end{aligned}
\end{equation}
and correspondingly 
\begin{equation}
\begin{aligned}
\Tilde{\gamma}=\frac{a_\sigma }{\pi b_t}=\frac{1}{2\pi}\sqrt{\frac{N+8}{\pi \epsilon}}+O\left(\epsilon^{1/2}\right).
    \end{aligned}
\end{equation}
From the above, we see that once the background dimension is set to $d=4-\epsilon$, the coupling $\tilde{\gamma}$ is determined and does not flow under RG. 

To perform the NL$\Sigma$M perturbative analysis, we introduce an IR regulation on the defect dimension $\tilde{d}$. We define $\tilde{d}\equiv2+\tilde{\epsilon}$ to distinguish it from the background dimension, and let $\tilde{\epsilon}\ll 1 $. In such a  dimensional regulation scheme, we also introduce a UV cutoff regulation denoted as $\Lambda$. From the viewpoint of defect modes, the effective action \eqref{IR effective} can be written as (after field redefinition 
and identifying $S_{\text{IR}}\equiv g\, S_{\text{IR}}^{\text{rescaled}}$):
\begin{equation}
\label{goldstone modes effective}
\begin{aligned}
S_{\text{IR}}^{\text{rescaled}}=&\int d^{2+\Tilde{\epsilon}}z\left\{\frac{1}{2 g}(\partial \pi)^2+\frac{1}{2\Lambda^{\Tilde{\epsilon}}}(\pi \partial \pi)^2\right.\\
&\left.-\frac{\Tilde{\gamma}^2}{2\Lambda^{\Tilde{\epsilon}}}\int d^{2+\Tilde{\epsilon}}{z}' \langle Q(z) Q({z}')\rangle_{\text{reg}}\pi(z)\cdot \pi({z}')+O\left(\Lambda^{-2\Tilde{\epsilon}}\right)\right\},
    \end{aligned}
\end{equation}

where the UV-divergence in the tilt two-point function $\langle Q(z) Q({z}')\rangle$ is regularized in such a way that preserves the  $O(M)$ symmetry. Here we slightly digress and explain how this is done: Tilt operators have a protected dimension $\Delta_Q=\tilde{d}$. \footnote{Note that here we are referring to the DCFT degrees of freedom.  The actions \eqref{goldstone modes effective}, \eqref{IR effective}, as will be elaborated on in what follows, exhibit a nontrivial RG flow. }
Hence the Fourier transformation of its two-point function yields: 
\begin{equation}
\label{Fourier Transformation}
\begin{aligned}
\int_{|z|\geq \frac{1}{\Lambda}}d^{\tilde{d}}z\frac{e^{-ikz}}{|z|^{2 \Tilde{d}}}=\frac{2 \pi
   ^{\frac{\tilde{d}}{2}} \Lambda ^{\tilde{d}}}{\tilde{d} \Gamma
   \left(\tilde{d}/2\right)}-\frac{k^2
   \pi ^{\frac{\tilde{d}}{2}} \Lambda ^{\tilde{d}-2}}{\left(\tilde{d}-2\right)
   \tilde{d} \Gamma \left(\tilde{d}/2\right)}+\frac{ \pi ^{\frac{\tilde{d}}{2}} k^{\tilde{d}} \Gamma
   \left(-\tilde{d}/2\right)}{2^{\tilde{d}}\Gamma \left(\tilde{d}\right)}+O\left(\Lambda ^{\tilde{d}-4}\right).
    \end{aligned}
\end{equation}
The non-local interaction term in \eqref{goldstone modes effective} is of the dimension $\Lambda^{2-\tilde{d}}$, so the first term in \eqref{Fourier Transformation} marks the UV divergence regulated by the cutoff $\Lambda$. Since the interaction is quadratic in $\{\pi^i\}$, such a term introduces a localized mass to the $\pi$-modes and explicitly breaks the $O(M)$ symmetry. In other words, the $\pi$-modes are required to be massless in order to preserve the $O(M)$ symmetry. We note that this resembles cases of ordinary spontaneous symmetry breaking, where Goldstone modes are massless. Therefore, the correlator followed by the required UV counterterm \cite{metlitski2022boundary,krishnan2023plane} as implied by the symmetry reads:
\begin{equation}
\begin{aligned}
\langle Q(z) Q({z}')\rangle_{\text{reg}}=\frac{1}{|z-{z}'|^{2 \Tilde{d}}}-\frac{\pi^{\Tilde{d}/2}\Lambda^{\Tilde{d}}}{\Gamma \left(\frac{\Tilde{d}}{2}+1\right)}\delta^{\Tilde{d}}(z-{z}').
    \end{aligned}
\end{equation}

A convenient way to study the NL$\Sigma$M RG flow is by investigating the (bare-)two-point vertex function $V^{(2)}_\pi(k)$ \cite{zinn2021quantum}. Working in momentum space, we find :
\begin{equation}
\begin{aligned}\label{eq_v2pi}
V^{(2)}_\pi(k)=k^2\Lambda^{\Tilde{\epsilon}}\left[\frac{1}{g}+\frac{1}{2 \pi  \tilde{\epsilon }}+\frac{\pi  \tilde{\gamma }^2}{2 \tilde{\epsilon }}+O\left(\Tilde{\epsilon}^0,\Lambda^{-\Tilde{\epsilon}}\right)\right].
    \end{aligned}
\end{equation}
In the above, the first term is the amputation of the free field propagator, the second is the one-loop contribution, and the third is the contribution of the non-local interaction in \eqref{goldstone modes effective}. It is a standard problem to extract the beta function from the renormalized vertex, and the details can be found in appendix \ref{Regulations in NLSM}. In the scheme we chose, the (minus-)beta function is:
\begin{equation}
\label{NLSM RG}
\begin{aligned}
-\beta(g)=-\Tilde{\epsilon} g+\left(\frac{(M-2)}{2\pi}-\frac{\pi}{2}(\Tilde{\gamma})^2\right)g^2+O\left(g^3\right),
    \end{aligned}
\end{equation}
which at the physical scenario of our interest $\tilde{d}=2$ reduces to:
\begin{equation}
\begin{aligned}
-\beta(g)=\left(\frac{(M-2)}{2\pi}-\frac{N+8}{8\pi^2\epsilon}+O\left(\epsilon^0\right)\right)g^2+O\left(g^3\right).
    \end{aligned}
\end{equation}

The above, therefore, implies that the non-locality in equation \eqref{goldstone modes effective} stabilizes $g=0$ as an IR fixed point\footnote{As shown in appendix \ref{Regulations in NLSM}, the field renormalization function also is fixed for $g=0$.}. Such a fact validates the action in equation \eqref{IR effective} as a possible IR effective action, in which $S_\sigma$ and $S_{\mathcal{D}}$ are weakly coupled. This provides the completion of the phase diagram from figure \ref{Phase Diagram}.
From the effective action \eqref{goldstone modes effective} one can show that the $\pi$-bosons acquire logarithmic correlation functions,  and the theory thus results in the extraordinary-log phase universality class, similar to the analysis in \cite{Nahum:2013xsa,metlitski2022boundary}. In particular, it implies that no spontaneous symmetry breaking takes place.

\acknowledgments
We thank G. Cuomo, L. Iliesiu, Z. Komargodski, L. Rastelli, and S. Shao for many useful discussions.  
We are  particularly grateful to G. Cuomo and Z. Komargodski for providing comments on a preliminary version of this manuscript. ARM is supported by the Simons Center for Geometry and Physics. 
ARM is an awardee of the Women’s Postdoctoral Career Development Award. 

\appendix 

\appendix 

\section{$\mathcal{F}$-function and Regulation}
\label{Regulation}
We first formally explain how equation \eqref{F-function} was obtained. Consider the coordinate system $x=\{r,\theta,\varphi,t\}$, and the $\mathbb{S}^2$ defect being inserted at $\{r=R,t=0\}$, such that

\begin{equation}
\begin{aligned}
S=\frac{1}{2}\int_{\mathbb{R}^{d}}d^{d}x\left\{(\partial \phi)^2+\delta(r-R)\delta^{1-\epsilon}(t)\gamma_{\text{b}}\phi^2\right\}.
    \end{aligned}
\end{equation}
One can perform the modes expansion:
\begin{equation}
\begin{aligned}
\phi(x)=\sum_{k>0}\sum_{l\geq |m|\geq0}\sum_{p} \sqrt{\frac{2}{\pi}}k j_l(kr)Y_l^m(\theta,\varphi) e^{ipt} \phi_{k,l,m,p}\,,
    \end{aligned}
\end{equation}
where $\{Y_l^m\}$ are the spherical harmonics and $\{j_l\}$ are the spherical Bessel functions. The Laplacian is diagonal with respect to such modes, and the action reads: 
\begin{equation}
\begin{aligned}
S=&\frac{1}{2}\sum_{l\geq |m|\geq0}\sum_{k_{1,2}>0}\sum_{p_{1,2}}\left\{(k_1^2+p_1^2)\delta_{k_1,k_2}\delta_{p_1+p_2}\right.\\ &\left.+\frac{2\gamma_{\text{b}} R^2}{\pi}k_1k_2j_l(k_1R)j_l(k_2R)\right\}\phi_{k_1,l,m,p_1}\phi_{k_2,l,m,p_2}\,.
    \end{aligned}
\end{equation}
Due to the defect's isometry, the action is diagonal in the $\{l,m\}$ mode indices and it will be enough to work out the determinant of the block matrix (denoted by $S_{l}$): 
\begin{equation} \label{eq_app4}
\begin{aligned}
\det\left( S_{l}\right)=&\det\left( k^2+p^2 \right) \left\{1+\gamma_{\text{b}} R^2\sum_{k,p}\frac{2k^2j_l(kR)^2}{\pi (k^2+p^2)}\right\}\\
=&\det\left( k^2+p^2 \right) \left[1+(\Lambda R)^{\epsilon}\alpha_l\right].
    \end{aligned}
\end{equation}
We recognize the second term in the above equation as the free field propagator between spherical harmonics and serve as a definition of \eqref{al def}. Now the $F$-function has the form as in \eqref{F-function}.

To derive equation \eqref{F-function IR limit}, it is useful to consider the binomial expansion
\begin{equation}
\begin{aligned}
(1-x)^{-\Tilde{d}}=\sum_{n\geq 1}\frac{(\Tilde{d})_n}{n!}x^n, \qquad \text{for } \Tilde{d}<0
    \end{aligned}
\end{equation}
Taking $x=1$ and analytically continuing $\Tilde{d}\rightarrow2$, one obtains:
\begin{equation}
\begin{aligned}
\sum_{l\geq 0}(2l+1)=0,
    \end{aligned}
\end{equation}
that is, equation \eqref{F-function} is independent of constant multiplicative factors in each $\log$-term, and the IR limit reduces to \eqref{F-function IR limit}. The summation \eqref{F-function IR limit} has been calculated in \cite{diaz2007partition}, and here we summarize the key steps.
Taking a derivative with respect to $\epsilon$ of both sides of  equation \eqref{F-function IR limit} reads:
\begin{equation}
\begin{aligned}
\frac{\partial \mathcal{F}}{\partial \epsilon}=\frac{\epsilon \sin{\left(\epsilon\pi/2\right) } }{4 \sin{(\Tilde{d}\pi/2) }}\frac{\Gamma \left(\frac{\Tilde{d}}{2}-\frac{\epsilon}{2}\right) \Gamma \left(\frac{\Tilde{d}}{2}+\frac{\epsilon}{2}\right)}{\Gamma (\Tilde{d}+1)},
    \end{aligned}
\end{equation}
which diverges for even integer $\Tilde{d}$. Expanding the defect dimension around the physical one of our interest $\Tilde{d}=2+\Tilde{\epsilon}$, we get:
\begin{equation}
\begin{aligned}\label{eq_temp}
\frac{\partial \mathcal{F}}{\partial \epsilon}=-\frac{\epsilon^2}{8\Tilde{\epsilon}}+O\left((\Tilde{\epsilon})^0\right),
    \end{aligned}
\end{equation}
such a pole-term is the Weyl anomaly in dimensional regularization scheme \cite{graham1999volume}. Notice at $\epsilon=0$ the defect is trivial and therefore $\mathcal{F}=0$ is free of Weyl anomaly. To obtain the $b$-coefficient, one can simply integrate \eqref{eq_temp} with respect to $\epsilon$. Note that the $1/3$ factor that arises from the integration is the $\mathbb{S}^2$ geometric factor, and is not a part of the definition of $b$, as can be seen from equation \eqref{b-def}.

\section{Minimal Subtraction}
\label{MS}

We elaborate the RG scheme we choose in section \ref{sec_TheON_critical_perturbationTheory}. The bare couplings with their corresponding counter terms are defined as 

\begin{equation}
\begin{aligned}
\gamma_b=&\Lambda^{\frac{6\epsilon}{N+8}}(\gamma+\frac{\delta \gamma}{\epsilon}),\\
(g_{ij})_b=&\Lambda^{\frac{N+6}{N+8}\epsilon}(g_{ij}+\frac{\delta g_{ij}}{\epsilon}).
    \end{aligned}
\end{equation}
To cancel the $O(1/\epsilon)$ divergence in the one-point functions \eqref{scalar one pt} and \eqref{stt one pt}, the counter terms are specified as:
\begin{equation}
\begin{aligned}
\delta \gamma=&\frac{N+8}{12 \pi  N }\gamma^2+\frac{(N+8)}{\pi  (N+3) }\text{Tr}\{g^2\},\\
\delta g_{ij}=&\frac{N+8}{12 \pi  N } g_{ij} \gamma+\frac{N+8}{2\pi (N+6)  }\left((g^2)_{ij}-\frac{\delta_{ij}}{N}\text{Tr}\{g^2\}\right).
    \end{aligned}
\end{equation}
At the trivial fixed point, the leading term in the beta functions can be read from the bulk scaling dimension of the corresponding conformal operator:

\begin{equation}
\begin{aligned}
-\beta(\gamma):=&-\Lambda\frac{\partial}{\partial \Lambda}\gamma=\frac{6\epsilon}{N+8}\gamma+O\left(\gamma^2,\gamma g,g^2\right ),\\
-\beta(g_{ij}):=&-\Lambda\frac{\partial}{\partial \Lambda}g_{ij}=\frac{N+6}{N+8}\epsilon g_{ij}+O\left(\gamma^2,\gamma g,g^2\right ). 
    \end{aligned}
\end{equation}
By requiring:

\begin{equation}
\begin{aligned}
\Lambda\frac{\partial}{\partial \Lambda}\gamma_b=\Lambda\frac{\partial}{\partial \Lambda}(g_{ij})_b=0,
    \end{aligned}
\end{equation}
one can obtain the beta functions \eqref{scalar beta} and \eqref{stt beta} in agreement with standard conformal perturbation theory formalism \cite{komargodski2017random}.

\section{RG in NL$\Sigma$M}
\label{Regulations in NLSM}
In this appendix, we elaborate on the NLSM RG flow details and explain how the beta-function \eqref{NLSM RG} was derived. At the one-loop level, we will use the following integral for the divergence of $\pi$-modes correlation at coincidence points:
\begin{equation}
\begin{aligned}
\int _{|k|\leq \Lambda}\frac{d^{\Tilde{d}}k}{(2 \pi)^{\Tilde{d}}} \frac{g}{k^2}=\frac{g \Lambda
   ^{\tilde{\epsilon }}}{2^{\tilde{\epsilon }} \pi ^{\frac{\tilde{\epsilon }}{2}+1} \tilde{\epsilon }^2 \Gamma \left(\frac{\tilde{\epsilon
   }}{2}\right)}=\frac{g\Lambda^{\Tilde{\epsilon}}}{2 \pi  \tilde{\epsilon }}+O\left(\tilde{\epsilon }^0\right).
    \end{aligned}
\end{equation}
Let us investigate the one-point function of $\langle (\vec{n})^1\rangle$. The $\pi$-modes in this case are dimensionful, and we introduce the sliding scale $\mu$ such that $(\Vec{n})^1=\sqrt{1-(\pi^i)^2/\mu^{\tilde{\epsilon}}}$. We find: 
\begin{equation}
\begin{aligned}
\langle (\vec{n})^1\rangle=1-\frac{\langle(\pi^i)^2\rangle}{2\mu^{\tilde{\epsilon}}}+O\left(g^2\right)=1-\frac{M-1}{4\pi \tilde{\epsilon}}g\left(\frac{\Lambda}{\mu}\right)^{\tilde{\epsilon }}+O\left(g^2\right).
    \end{aligned}
\end{equation}
The divergence in $\Lambda$ in the above expression should be cured by a field renormalization $Z_n$, defined by $\Vec{n}\equiv Z_n^{1/2}\Vec{n}_r$, where $\Vec{n}_r$ is the renormalized field, such that the `anomalous dimension' $\eta_n$ \cite{zinn2021quantum,henriksson2023critical,metlitski2022boundary} is defined as:
\begin{equation}
\begin{aligned}
Z_n=&1-\frac{M-1}{2\pi\tilde{\epsilon}}g\left(\frac{\Lambda}{\mu}\right)^{\tilde{\epsilon }}+O\left(g^2\right),\\
\eta_n:=&\mu\frac{\partial}{\partial\mu}\log{Z_n}=\frac{M-1}{2\pi}g+O\left(\tilde{\epsilon} g, g^2\right).
    \end{aligned}
\end{equation}
Note that from the above result, we see that $\eta_n$ indeed vanishes when $g=0$. Next, we require the renormalized two-point vertex to be independent of the cutoff $\Lambda$, where we define the renormalized correlator:
\begin{equation}
\begin{aligned}
\left(V^{(2)}_\pi\right)_{\text{r}}=Z_nV^{(2)}_\pi,
    \end{aligned}
\end{equation}
and $V^{(2)}_\pi$ is given by equation \eqref{eq_v2pi}. The Callan–Symanzik equation therefore reads:
\begin{equation}
\begin{aligned}
\left(\Lambda \frac{\partial}{\partial\Lambda} +\beta(g)\frac{\partial}{\partial g}-\eta_n\right)V^{(2)}_\pi=0\,,
    \end{aligned}
\end{equation}
and from it, one easily obtains the result \eqref{NLSM RG}.

\bibliography{ref}

\providecommand{\href}[2]{#2}\begingroup\raggedright\begin{thebibliography}{10}

\bibitem{gaiotto2015generalized}
D.~Gaiotto, A.~Kapustin, N.~Seiberg, and B.~Willett, {\it Generalized global
  symmetries},  {\em Journal of High Energy Physics} {\bf 2015} (2015), no.~2
  1--62.

\bibitem{roumpedakis2022higher}
K.~Roumpedakis, S.~Seifnashri, and S.-H. Shao, {\it Higher gauging and
  non-invertible condensation defects},  {\em arXiv preprint arXiv:2204.02407}
  (2022).

\bibitem{aharony2022phases}
O.~Aharony, G.~Cuomo, Z.~Komargodski, M.~Mezei, and A.~Raviv-Moshe, {\it Phases
  of wilson lines in conformal field theories},  {\em arXiv preprint
  arXiv:2211.11775} (2022).

\bibitem{gukov2006gauge}
S.~Gukov and E.~Witten, {\it Gauge theory, ramification, and the geometric
  langlands program},  {\em arXiv preprint hep-th/0612073} {\bf 104} (2006).

\bibitem{drukker2008probing}
N.~Drukker, J.~Gomis, and S.~Matsuura, {\it Probing $n= 4$ sym with surface
  operators},  {\em Journal of High Energy Physics} {\bf 2008} (2008), no.~10
  048.

\bibitem{gukov2010rigid}
S.~Gukov and E.~Witten, {\it Rigid surface operators},  {\em Advances in
  Theoretical and Mathematical Physics} {\bf 14} (2010), no.~1 87--178.

\bibitem{Wang:2020seq}
Y.~Wang, {\it {Taming defects in $ \mathcal{N} $ = 4 super-Yang-Mills}},  {\em
  JHEP} {\bf 08} (2020), no.~08 021,
  [\href{http://arxiv.org/abs/2003.11016}{{\tt arXiv:2003.11016}}].

\bibitem{Herzog:2022jqv}
C.~P. Herzog and A.~Shrestha, {\it {Conformal surface defects in Maxwell theory
  are trivial}},  {\em JHEP} {\bf 08} (2022) 282,
  [\href{http://arxiv.org/abs/2202.09180}{{\tt arXiv:2202.09180}}].

\bibitem{lauria2021line}
E.~Lauria, P.~Liendo, B.~C. van Rees, and X.~Zhao, {\it Line and surface
  defects for the free scalar field},  {\em Journal of High Energy Physics}
  {\bf 2021} (2021), no.~1 1--36.

\bibitem{krishnan2023plane}
A.~Krishnan and M.~A. Metlitski, {\it A plane defect in the 3d o $(n) $ model},
   {\em arXiv preprint arXiv:2301.05728} (2023).

\bibitem{burkhardt1987surface}
T.~Burkhardt and J.~Cardy, {\it Surface critical behaviour and local operators
  with boundary-induced critical profiles},  {\em Journal of Physics A:
  Mathematical and General} {\bf 20} (1987), no.~4 L233.

\bibitem{mcavity1995conformal}
D.~M. McAvity and H.~Osborn, {\it Conformal field theories near a boundary in
  general dimensions},  {\em Nuclear Physics B} {\bf 455} (1995), no.~3
  522--576.

\bibitem{ohno19831}
K.~Ohno and Y.~Okabe, {\it The 1/n expansion for the n-vector model in the
  semi-infinite space},  {\em Progress of theoretical physics} {\bf 70} (1983),
  no.~5 1226--1239.

\bibitem{dimofte2018dual}
T.~Dimofte, D.~Gaiotto, and N.~M. Paquette, {\it Dual boundary conditions in 3d
  scft’s},  {\em Journal of High Energy Physics} {\bf 2018} (2018), no.~5
  1--101.

\bibitem{metlitski2022boundary}
M.~Metlitski, {\it Boundary criticality of the o (n) model in d= 3 critically
  revisited},  {\em SciPost Physics} {\bf 12} (2022), no.~4 131.

\bibitem{Cuomo:2022xgw}
G.~Cuomo, Z.~Komargodski, M.~Mezei, and A.~Raviv-Moshe, {\it {Spin impurities,
  Wilson lines and semiclassics}},  {\em JHEP} {\bf 06} (2022) 112,
  [\href{http://arxiv.org/abs/2202.00040}{{\tt arXiv:2202.00040}}].

\bibitem{jensen2016constraint}
K.~Jensen and A.~O’Bannon, {\it Constraint on defect and boundary
  renormalization group flows},  {\em Physical Review Letters} {\bf 116}
  (2016), no.~9 091601.

\bibitem{wang2021surface}
Y.~Wang, {\it Surface defect, anomalies and b-extremization},  {\em Journal of
  High Energy Physics} {\bf 2021} (2021), no.~11 1--33.

\bibitem{shachar2022rg}
T.~Shachar, R.~Sinha, and M.~Smolkin, {\it Rg flows on two-dimensional
  spherical defects},  {\em arXiv preprint arXiv:2212.08081} (2022).

\bibitem{henningson1999weyl}
M.~Henningson and K.~Skenderis, {\it Weyl anomaly for wilson surfaces},  {\em
  Journal of High Energy Physics} {\bf 1999} (1999), no.~06 012.

\bibitem{schwimmer2008entanglement}
A.~Schwimmer and S.~Theisen, {\it Entanglement entropy, trace anomalies and
  holography},  {\em Nuclear physics B} {\bf 801} (2008), no.~1-2 1--24.

\bibitem{graham1999conformal}
C.~R. Graham and E.~Witten, {\it Conformal anomaly of submanifold observables
  in ads/cft correspondence},  {\em Nuclear Physics B} {\bf 546} (1999),
  no.~1-2 52--64.

\bibitem{PhysRevE.72.016128}
Y.~Deng, H.~W.~J. Bl\"ote, and M.~P. Nightingale, {\it Surface and bulk
  transitions in three-dimensional $\mathrm{O}(n)$ models},  {\em Phys. Rev. E}
  {\bf 72} (Jul, 2005) 016128.

\bibitem{PhysRevE.73.056116}
Y.~Deng, {\it Bulk and surface phase transitions in the three-dimensional
  $o(4)$ spin model},  {\em Phys. Rev. E} {\bf 73} (May, 2006) 056116.

\bibitem{ParisenToldin:2020gpb}
F.~Parisen~Toldin, {\it {Boundary Critical Behavior of the Three-Dimensional
  Heisenberg Universality Class}},  {\em Phys. Rev. Lett.} {\bf 126} (2021),
  no.~13 135701, [\href{http://arxiv.org/abs/2012.00039}{{\tt
  arXiv:2012.00039}}].

\bibitem{Hu:2021xdy}
M.~Hu, Y.~Deng, and J.-P. Lv, {\it {Extraordinary-Log Surface Phase Transition
  in the Three-Dimensional XY Model}},  {\em Phys. Rev. Lett.} {\bf 127}
  (2021), no.~12 120603, [\href{http://arxiv.org/abs/2104.05152}{{\tt
  arXiv:2104.05152}}].

\bibitem{Toldin:2021kun}
F.~P. Toldin and M.~A. Metlitski, {\it {Boundary Criticality of the 3D O(N)
  Model: From Normal to Extraordinary}},  {\em Phys. Rev. Lett.} {\bf 128}
  (2022), no.~21 215701, [\href{http://arxiv.org/abs/2111.03613}{{\tt
  arXiv:2111.03613}}].

\bibitem{Padayasi:2021sik}
J.~Padayasi, A.~Krishnan, M.~A. Metlitski, I.~A. Gruzberg, and M.~Meineri, {\it
  {The extraordinary boundary transition in the 3d O(N) model via conformal
  bootstrap}},  {\em SciPost Phys.} {\bf 12} (2022), no.~6 190,
  [\href{http://arxiv.org/abs/2111.03071}{{\tt arXiv:2111.03071}}].

\bibitem{2023Giombi}
S.~Giombi and B.~Liu, {\it {Notes on a Surface Defect in the $O(N)$ Model}},
  \href{http://arxiv.org/abs/2305.11402}{{\tt arXiv:2305.11402}}.

\bibitem{Trepanier:2023tvb}
M.~Tr\'epanier, {\it {Surface defects in the $O(N)$ model}},
  \href{http://arxiv.org/abs/2305.10486}{{\tt arXiv:2305.10486}}.

\bibitem{wilson1974renormalization}
K.~G. Wilson and J.~Kogut, {\it The renormalization group and the $\varepsilon$
  expansion},  {\em Physics reports} {\bf 12} (1974), no.~2 75--199.

\bibitem{cuomo2022localized}
G.~Cuomo, Z.~Komargodski, and M.~Mezei, {\it Localized magnetic field in the o
  (n) model},  {\em Journal of High Energy Physics} {\bf 2022} (2022), no.~2
  1--50.

\bibitem{Cuomo:2021rkm}
G.~Cuomo, Z.~Komargodski, and A.~Raviv-Moshe, {\it {Renormalization Group Flows
  on Line Defects}},  {\em Phys. Rev. Lett.} {\bf 128} (2022), no.~2 021603,
  [\href{http://arxiv.org/abs/2108.01117}{{\tt arXiv:2108.01117}}].

\bibitem{gubser2003universal}
S.~S. Gubser and I.~R. Klebanov, {\it A universal result on central charges in
  the presence of double-trace deformations},  {\em Nuclear Physics B} {\bf
  656} (2003), no.~1-2 23--36.

\bibitem{klebanov2011f}
I.~R. Klebanov, S.~S. Pufu, and B.~R. Safdi, {\it F-theorem without
  supersymmetry},  {\em Journal of High Energy Physics} {\bf 2011} (2011),
  no.~10 1--27.

\bibitem{diaz2007partition}
D.~E. Diaz and H.~Dorn, {\it Partition functions and double-trace deformations
  in ads/cft},  {\em Journal of High Energy Physics} {\bf 2007} (2007), no.~05
  046.

\bibitem{graham1999volume}
C.~R. Graham, {\it Volume and area renormalizations for conformally compact
  einstein metrics},  {\em arXiv preprint math/9909042} (1999).

\bibitem{dey2020operator}
P.~Dey, T.~Hansen, and M.~Shpot, {\it Operator expansions, layer susceptibility
  and two-point functions in bcft},  {\em Journal of High Energy Physics} {\bf
  2020} (2020), no.~12 1--34.

\bibitem{shpot2021boundary}
M.~Shpot, {\it Boundary conformal field theory at the extraordinary transition:
  The layer susceptibility to o ($\varepsilon$)},  {\em Journal of High Energy
  Physics} {\bf 2021} (2021), no.~1 1--28.

\bibitem{ginsparg1988applied}
P.~Ginsparg, {\it Applied conformal field theory},  {\em arXiv preprint
  hep-th/9108028} (1988).

\bibitem{kehrein1995spectrum}
S.~K. Kehrein, {\it The spectrum of critical exponents in ($\phi$2) 2 theory in
  d= 4- $\varepsilon$ dimensions resolution of degeneracies and hierarchical
  structures},  {\em Nuclear Physics B} {\bf 453} (1995), no.~3 777--806.

\bibitem{henriksson2023critical}
J.~Henriksson, {\it The critical o (n) cft: Methods and conformal data},  {\em
  Physics Reports} {\bf 1002} (2023) 1--72.

\bibitem{komargodski2017random}
Z.~Komargodski and D.~Simmons-Duffin, {\it The random-bond ising model in 2.01
  and 3 dimensions},  {\em Journal of Physics A: Mathematical and Theoretical}
  {\bf 50} (2017), no.~15 154001.

\bibitem{Billo:2016cpy}
M.~Bill\`o, V.~Gon\c{c}alves, E.~Lauria, and M.~Meineri, {\it {Defects in
  conformal field theory}},  {\em JHEP} {\bf 04} (2016) 091,
  [\href{http://arxiv.org/abs/1601.02883}{{\tt arXiv:1601.02883}}].

\bibitem{Cuomo:2021cnb}
G.~Cuomo, M.~Mezei, and A.~Raviv-Moshe, {\it {Boundary conformal field theory
  at large charge}},  {\em JHEP} {\bf 10} (2021) 143,
  [\href{http://arxiv.org/abs/2108.06579}{{\tt arXiv:2108.06579}}].

\bibitem{zinn2021quantum}
J.~Zinn-Justin, {\em Quantum field theory and critical phenomena}, vol.~171.
\newblock Oxford university press, 2021.

\bibitem{2023Cuomo}
G.~Cuomo and S.~Zhang, {\it Spontaneous symmetry breaking on surface defects},
  {\em to appear.}

\bibitem{nishioka2021free}
T.~Nishioka and Y.~Sato, {\it Free energy and defect c-theorem in free scalar
  theory},  {\em Journal of High Energy Physics} {\bf 2021} (2021), no.~5
  1--56.

\bibitem{d2004supersymmetric}
E.~D’Hoker and D.~Z. Freedman, {\it Supersymmetric gauge theories and the
  ads/cft correspondence},  in {\em Strings, Branes and Extra Dimensions: TASI
  2001}, pp.~3--159.
\newblock World Scientific, 2004.

\bibitem{liendo2013bootstrap}
P.~Liendo, L.~Rastelli, and B.~C. van Rees, {\it The bootstrap program for
  boundary $cft_d$},  {\em Journal of High Energy Physics} {\bf 2013} (2013),
  no.~7 1--52.

\bibitem{gaiotto2014bootstrapping}
D.~Gaiotto, D.~Mazac, and M.~F. Paulos, {\it Bootstrapping the 3d ising twist
  defect},  {\em Journal of High Energy Physics} {\bf 2014} (2014), no.~3
  1--34.

\bibitem{Nahum:2013xsa}
A.~Nahum, P.~Serna, A.~M. Somoza, and M.~Ortu\~no, {\it {Loop models with
  crossings}},  {\em Phys. Rev. B} {\bf 87} (2013), no.~18 184204,
  [\href{http://arxiv.org/abs/1303.2342}{{\tt arXiv:1303.2342}}].

\end{thebibliography}\endgroup
	\bibliographystyle{JHEP.bst}

\end{document}